\documentclass[11pt]{article}

\usepackage{amsmath}
\usepackage{graphicx}
\usepackage{indentfirst}
\usepackage{amssymb}
\usepackage{cite}
\usepackage{color}
\usepackage{subfigure}

\begin{document}

\title{\bf{Weak Cosmic Censorship Conjecture\\in Kerr-Newman-(Anti-)de Sitter Black Hole\\
with Charged Scalar Field}}

\date{}
\maketitle

\begin{center}
\author{Bogeun Gwak}$^a$\footnote{rasenis@dgu.ac.kr}\\

\vskip 0.25in
$^{a}$\it{Division of Physics and Semiconductor Science, Dongguk University, Seoul 04620,\\Republic of Korea}\\
\end{center}
\vskip 0.6in

{\abstract
{We investigate the weak cosmic censorship conjecture in extremal and near-extremal Kerr-Newman-(anti-)de Sitter black holes by the scattering of a massive scalar field with an electric charge. Under this scattering, the scalar field fluxes change the black hole state, as determined by the mass, angular momentum, and electric charge. The black hole may exceed its extremal condition because of these changes. However, we find that the black hole cannot be overcharged or overspun by the scattering. In particular, although the fluxes are closely associated with the asymptotic boundary conditions along the flat, anti-de Sitter, and de Sitter spacetimes, the weak cosmic censorship conjecture is valid for any scalar field boundary conditions. Moreover, the validity of the weak cosmic censorship conjecture is thermodynamically preferred for this scattering.}}

\thispagestyle{empty}
\newpage
\setcounter{page}{1}

\section{Introduction}\label{sec1}

Black holes were theoretically predicted in general relativity as the end states of collapsing massive stars\cite{Penrose:1969pc}. With recent technological developments and the detection of gravitational waves from binary black holes\cite{TheLIGOScientific:2016src}, black holes have become viable objects\cite{Akiyama:2019cqa,Akiyama:2019brx}. Moreover, black holes have become highly interesting research topics as a result of these developments. Black holes consist of an event horizon and a singularity. The event horizon is a spherical surface with no outgoing geodesics. Inside the horizon, an ingoing geodesic moves to the center of the black hole and ends at the singularity. However, owing to the horizon, an outside observer cannot view the singularity or the interior of a black hole\cite{Penrose:1964wq}. Classically, the mass of a black hole is divided into irreducible mass and reducible energy. The irreducible mass never decreases. Thus, the reducible energy is only changeable via an interaction\cite{Christodoulou:1970wf,Christodoulou:1972kt}. When the mass of a black hole decreases through the Penrose process\cite{Penrose:1971uk}, the decrease in the reducible component exceeds the increase in the irreducible component. The specifics of this scenario vary according to the associated quantum physics. A marginal amount of energy, called Hawking radiation, can be emitted by black holes \cite{Hawking:1974sw,Hawking:1976de}. On account of this radiation, black holes can be treated as thermal objects with Hawking temperature. Furthermore, the irreducible mass behaves as an entropy. Thus, the surface area of the black hole, which is the square of its irreducible mass, is associated with the entropy as a thermodynamic conjugate variable to Hawking temperature. This is called the Bekenstein-Hawking entropy\cite{Bekenstein:1973ur,Bekenstein:1974ax}. Hence, black hole thermodynamics can be suitably defined in terms of the black hole variables under the laws of thermodynamics.

The event horizon of a black hole encompasses its curvature singularity and  plays a significant role in preventing measurement of the singularity by an outside observer. Without the horizon, an emission from a naked singularity would reach an outside observer and cause a breakdown of causality in the spacetime. To preclude such a breakdown, the weak cosmic censorship (WCC) conjecture predicts that no naked singularity can be measured by a static observer outside a black hole\cite{Penrose:1964wq,Penrose:1969pc}. In fact, the Schwarzschild black hole is free from such a breakdown, because the horizon never disappears about a Schwarzschild black hole of any mass. However, with the inclusion of more than one conserved quantity, most black holes have an extremal condition for the minimum mass, under which the black hole solution no longer has a horizon. If a naked singularity appears because of the extremal condition being exceeded, the WCC conjecture becomes invalid. Thus, the proof of the WCC conjecture focuses on whether a black hole can violate the extremal condition through overcharging or overspinning because, under the extremal condition, the horizon prevents measurement of the singularity. However, there is no general proof of the WCC conjecture. Hence, this conjecture must be tested for each particular case. The first investigation of the WCC conjecture was performed for an extremal Kerr black hole considering the effects of a point particle\cite{Wald:1974ge}. It was found that the black hole was never overspun, even if it received angular momentum from the particle. Hence, the WCC conjecture was validated. Without a general proof, the WCC conjecture depends on the approaches used to test the black hole. For a Reissner-Nordstr\"{o}m (RN) black hole, one test for the WCC conjecture indicated that the effects of the backreaction are negligible\cite{Hubeny:1998ga}. However, a detailed analysis of the WCC conjecture for this case showed that the backreaction between the RN black hole and a particle is considerable, and that the conjecture is valid with this consideration\cite{Isoyama:2011ea}. Furthermore, in WCC conjecture tests, the initial state of the black hole is crucial. Even though the WCC conjecture is suitable for an extremal Kerr black hole, the opposite conclusion has been reached for the near-extremal Kerr black hole\cite{Jacobson:2009kt}. These contradictions are resolved by considering the self-force\cite{Barausse:2010ka,Barausse:2011vx,Colleoni:2015ena,Colleoni:2015afa,Sorce:2017dst}. Overall, because the WCC conjecture is associated with many black hole effects under various theories, it is still tested from various perspectives \cite{BouhmadiLopez:2010vc,Gwak:2011rp,Rocha:2011wp,Gao:2012ca,Hod:2013vj,Zhang:2013tba,Rocha:2014jma,Gwak:2015fsa,Cardoso:2015xtj,Siahaan:2015ljs,Horowitz:2016ezu,Duztas:2016xfg,Gwak:2017kkt,Revelar:2017sem,Song:2017mdx,Liang:2018wzd,Yu:2018eqq,McInnes:2019rtj,Zeng:2019jrh,Wang:2019jzz,He:2019kws,Hu:2019zxr,Wang:2020osg,Jiang:2020fgr,Shaymatov:2020wtj,Ying:2020bch,Ahmed:2020ksf,Khodabakhshi:2020fkd}. Moreover, instead of tests based on a particle, the WCC conjecture can also be studied using probe fields as scattering processes \cite{Hod:2008zza,Semiz:2005gs,Toth:2011ab,Duztas:2013wua,Semiz:2015pna,Natario:2016bay,Duztas:2017lxk,Gwak:2018akg,Chen:2019nsr,Natario:2019iex,Jiang:2019vww,Wang:2019bml,Gwak:2019asi,Yang:2020iat,Hong:2020zcf,Feng:2020tyc,Yang:2020czk,Gwak:2020zht,Goncalves:2020ccm}. In investigations of the WCC conjecture, the test matter changes the black holes along with their conserved quantities. Such changes are then written as a dispersion relation, which corresponds to the first law of thermodynamics\cite{Gwak:2018akg}. Moreover, the Bekenstein-Hawking entropy never decreases in response to the test matter\cite{Gwak:2019rcz}, as ensured by the second law of thermodynamics. Finally, as regards the Hawking temperature, the zero-temperature state or extremal black hole is not reached through such finite operations\cite{Gwak:2020zht}. These findings imply that the classical tests of the WCC conjecture are closely associated with the related quantum physics through the laws of thermodynamics. Thus, the quantum effects may play an important role in elucidating and proving the WCC conjecture for various black-hole spacetimes.

In black holes, the cosmological constant is closely associated with the geometry of the asymptotic boundary. For a negative cosmological constant, the asymptotic boundary is an anti-de Sitter (AdS) spacetime. One of the most important advances concerning the AdS spacetime is the anti-de Sitter/conformal field theory (AdS/CFT) correspondence\cite{Maldacena:1997re,Gubser:1998bc,Witten:1998qj,Aharony:1999ti}, which ensures that the gravity theory defined in the $D$-dimensional AdS spacetime is associated with the CFT defined on its $(D-1)$-dimensional AdS boundary. The AdS black hole plays an important role in imposing the temperature on the dual CFT on the boundary. An AdS spacetime with no black hole is dual to the zero-temperature CFT. Furthermore, for an AdS black hole, the dual CFT is at a finite temperature that corresponds to the Hawking temperature of the black hole\cite{Witten:1998zw}. Notably, AdS/CFT correspondence has been applied to quantum chromodynamics and condensed matter theory. Another interesting phenomenon associated with the AdS spacetime is the superradiant instability of the black hole. For superradiance, the black-hole energy is extracted via a scattering process\cite{Zeldovich:1971aa,Zeldovich:1972ab}. As the AdS boundary is reflective, the scattering waves can be amplified and, eventually, the black hole becomes unstable. This instability is found in various AdS black holes\cite{Hawking:1999dp,Cardoso:2004hs,Cardoso:2006wa,Uchikata:2009zz}. Details of the superradiance are provided in \cite{Brito:2015oca} and the references therein. For a positive cosmological constant, the observable space is confined by the cosmological horizon, which is physically similar to the outer horizon of a black hole. In the case of a black hole with de Sitter (dS) geometry, the static observer is between the outer and cosmological horizons only. For a wave, the cosmological horizon acts as an absorbing boundary, and thus, particles or waves dissipate into the horizon. Various studies have been conducted on wave scattering in such dS black holes \cite{Teukolsky:1973ha,Suzuki:1998vy,Suzuki:1999nn,Cho:2009wf,Yoshida:2010zzb,Dias:2013sdc,Cardoso:2013pza,Zhang:2014kna,Delice:2015zga,Ganchev:2016zag,Crispino:2013pya,Kanti:2014dxa,Pappas:2016ovo,Cuyubamba:2016cug}. Recently, the quasinormal modes (QNMs) of dS black holes were extensively studied considering the strong cosmic censorship conjecture\cite{Cardoso:2017soq,Burikham:2017gdm,Gwak:2018rba,Hod:2018dpx,Dias:2018etb,Destounis:2020yav,Mishra:2019ged}. Furthermore, the QNMs of black holes itself can be useful for analysis of the ringdown phase of the binary merger and its final state \cite{Kokkotas:1999bd,Ferrari:2007dd,Berti:2009kk,Kamaretsos:2011um,Kamaretsos:2012bs}.

In this work, we investigate the validity of the WCC conjecture for near-extremal Kerr-Newman-(anti-)de Sitter (KN(A)dS) black holes by considering the scattering of a massive scalar field with an electric charge. KN(A)dS black holes are electrically charged and rotating solutions. Thus, the {\it{energy, angular momentum, and electric charge}} carried by the scalar field can perturb the corresponding conserved quantities of the black hole. This is the most {\it{general case}} of the Einstein-Maxwell theory of gravity in four dimensions. The cosmological constant is herein considered arbitrary, and we explore all possible boundaries of the {\it{flat, AdS}}, and {\it{dS spacetimes}}. In particular, since we test the WCC conjecture using a scalar field, the changes in the black hole depend on the scalar field profile, which must be sensitive to the asymptotic geometries determining the boundary conditions. We focus on whether the validity of the WCC conjecture depends on the boundary condition. Although the WCC conjecture has been studied extensively, its relationship to the boundary condition remains unclear. Our investigation proves that the WCC conjecture is {\it{valid for any boundary condition}} of the scalar field. Furthermore, we test both {\it{Kerr-type near-extremal}} and {\it{extremal}} KN(A)dS black holes to validate the WCC conjecture, as well as {\it{Nariai-type near-extremal}} black holes as an extension. Although this extension is not associated with the WCC conjecture, it shows the {\it{instability}} of a Nariai-type near-extremal black hole. Finally, the scalar field energy is assumed to be infinitesimally small. Therefore, our investigation is based on the first-order variation along the scalar field. As most individual pieces of matter are rather small compared with a black hole, this is physically reasonable. As we consider the first order, our analysis is {\it{consistent}} with the laws of thermodynamics, which is shown to be true at all boundaries. Ultimately, we show that {\it{the validity of the WCC conjecture is thermodynamically preferred}} under the scattering of the scalar field. This firmly establishes the validity of the WCC conjecture.

The remainder of this paper is organized as follows. In Section\,\ref{sec2}, we review KN(A)dS black holes and introduce coordinate transformations to obtain the correct thermodynamic variables. In Section\,\ref{sec3}, the scalar field solutions are found to be of the Schr\"{o}dinger-type equation. In Section\,\ref{sec4}, the conserved quantities carried by the scalar field are demonstrated in terms of their fluxes. In Section\,\ref{sec5}, the WCC conjecture is investigated using arbitrary cosmological constants. The general procedure is presented for the flat case. Then, the AdS and dS cases are tested and generalized to the integrated view. We also consider a Nariai-type near-extremal black hole. In Section\,\ref{sec6}, we relate our investigations of the WCC conjecture with the laws of thermodynamics. In Section\,\ref{sec7}, we summarize our conclusions.

\section{Kerr-Newman-(Anti-)de Sitter Black Holes}\label{sec2}

We consider Kerr-Newman black holes with an arbitrary cosmological constant $\Lambda$. Thus, the cosmological constant can be zero, negative, or positive, with the sign determining the asymptotic geometry, namely, flat, AdS, or dS, respectively. The KN(A)dS metric is expressed in Boyer-Lindquist coordinates.
\begin{align}\label{eq:metric01}
ds^2&=-\frac{\Delta_r}{\rho^2}\left(dt-\frac{a\sin^2\theta}{\Xi} d\phi\right)^2+\frac{\rho^2}{\Delta_r}dr^2+\frac{\rho^2}{\Delta_\theta}d\theta^2+\frac{\Delta_\theta\sin^2\theta}{\rho^2}\left(a\,dt-\frac{r^2+a^2}{\Xi}d\phi\right)^2,\\
\Delta_r&=(r^2+a^2)(1-\frac{1}{3}\Lambda r^2)-2Mr+Q^2,\,\,\Delta_\theta=1+\frac{1}{3}\Lambda a^2 \cos^2\theta,\,\,\rho^2=r^2+a^2\cos^2\theta,\,\,\Xi=1+\frac{1}{3}\Lambda a^2,\nonumber
\end{align}
where the mass, spin, and electric parameters are denoted as $M$, $a$, and $Q$, respectively. These parameters are associated with the mass $M_\text{B}$, angular momentum $J_\text{B}$, and electric charge $Q_\text{B}$ of the black hole. The gauge field coupled with the black hole is\cite{Caldarelli:1999xj,Chen:2010jj}
\begin{align}
A=-\frac{Qr}{\rho^2}\left(dt-\frac{a \sin^2\theta}{\Xi}d\phi\right).
\end{align}
Each point of the spacetime has an angular velocity owing to the rotation of the black hole. When the cosmological constant is zero, the asymptotic geometry of Eq.\,(\ref{eq:metric01}) is the Minkowski spacetime with zero angular velocity. Thus, it is static. However, for a nonzero cosmological constant, the angular velocity is nonzero at the asymptotic limit, which implies that the asymptotic observer is not static. This causes difficulties in defining conserved quantities in such spacetimes and affects the thermodynamics and WCC conjecture\cite{Caldarelli:1999xj,Gwak:2015fsa}. However, this problem can be resolved through coordinate transformations to recover an asymptotically static observer with zero angular velocity\cite{Hawking:1998kw}.
\begin{align}\label{eq:transformation01}
t\rightarrow T,\quad \phi\rightarrow \Phi + \frac{1}{3}a\Lambda T.
\end{align}
Therefore, we adopt the coordinate system $(T,r,\theta,\Phi)$. From Eq.\,(\ref{eq:metric01}), the transformed metric of the KN(A)dS black hole is
\begin{align}\label{eq:transfmetrc01}
ds^2=-\frac{\Delta_r}{\rho^2\Xi^2}\left(\Delta_\theta dT-a\sin^2\theta d\Phi\right)^2+\frac{\rho^2}{\Delta_r}dr^2+\frac{\rho^2}{\Delta_\theta}d\theta^2+\frac{\Delta_\theta\sin^2\theta}{\rho^2\Xi^2}\left(a\left(1-\frac{1}{3}\Lambda r^2\right)dT-(r^2+a^2)d\Phi\right)^2.
\end{align}
As the transformations in Eq.\,(\ref{eq:transformation01}) constitute changes in time and the azimuthal angle, the gauge field is also transformed to
\begin{align}
A=-\frac{Qr}{\rho^2 \Xi}\left(\Delta_\theta dT-a\sin^2\theta d\Phi\right).
\end{align}
The mass, angular momentum, and electric charge of the black hole are \cite{Caldarelli:1999xj}
\begin{align}\label{eq:blackholemass01}
M_\text{B}=\frac{M}{\Xi^2},\quad J_\text{B}=\frac{Ma}{\Xi^2},\quad Q_\text{B}=\frac{Q}{\Xi},
\end{align}
which are also associated with the cosmological constant. At the outer horizon $r_\text{h}$, the Hawking temperature $T_\text{h}$ and Bekenstein-Hawking entropy of the black hole are defined, and the angular velocity and the electric potential are also found.
\begin{align}\label{eq:thermodynamicvar01}
T_\text{h}=\frac{r_\text{h}\left(1-\frac{\Lambda a^2}{3}-\frac{a^2+Q^2}{r_\text{h}^2}-\Lambda r_\text{h}^2\right)}{4\pi\left(r_\text{h}^2+a^2\right)},\quad S_\text{h}=\frac{1}{4}A_\text{h}=\frac{\pi(r_\text{h}^2+a^2)}{\Xi},\quad \Omega_\text{h}=\frac{a\left(1-\frac{1}{3}\Lambda r_\text{h}^2\right)}{r_\text{h}^2+a^2},\quad \Phi_\text{h}=\frac{r_\text{h}Q}{r_\text{h}^2+a^2}.
\end{align}
Using Eqs.\,(\ref{eq:blackholemass01}) and (\ref{eq:thermodynamicvar01}), the first law of thermodynamics is defined as
\begin{align}\label{eq:1stlaw01}
d M_\text{B} = T_\text{h} d S_\text{h}+\Omega_\text{h}d J_\text{B} +\Phi_\text{h}dQ_\text{B}.
\end{align}
In this study, we independently find the first law from the change in the mass of the black hole without using Eq.\,(\ref{eq:1stlaw01}). Furthermore, the characteristics of the black hole depend on the sign of the cosmological constant. For a positive cosmological constant, another horizon, called the cosmological horizon, exists outside the black hole. At the cosmological horizon $r_\text{c}$, an angular velocity and electric potential are also given as 
\begin{align}\label{eq:thermodynamicvar05}
\Omega_\text{c}=\frac{a\left(1-\frac{1}{3}\Lambda r_\text{c}^2\right)}{r_\text{c}^2+a^2},\quad \Phi_\text{c}=\frac{r_\text{c}Q}{r_\text{c}^2+a^2},
\end{align}
These are significant for a Nariai black hole and we discuss them in the following sections.

\section{Solution to Scalar Field Equation}\label{sec3}

We herein consider the scattering of a massive scalar field with an electric charge. The scattering affects the state of the KN(A)dS black hole by carrying conserved quantities. To investigate the details of the changes in state, we must solve the scalar field equation. A scalar field solution is specified by two boundary conditions, one at the outer horizon and the other in the spatially asymptotic region. Thus, we focus on the relationship between the black hole and the two aforementioned boundary conditions. 

The action of the charged scalar field with a covariant derivative is expressed as
\begin{align}
S_\Psi =-\frac{1}{2}\int d^4 x \sqrt{-g}\left(\mathcal{D}_\mu \Psi {\mathcal{D}^*}^\mu \Psi^*+\mu^2\Psi\Psi^*\right),\quad \mathcal{D}_\mu = \partial_\mu - iqA_\mu,
\end{align}
where the scalar field mass and electric charge are denoted by $\mu$ and $q$, respectively. Then, the scalar field equations are
\begin{align}\label{eq:fieldequation01}
\frac{1}{\sqrt{-g}} \mathcal{D}_\mu \left(\sqrt{-g} g^{\mu\nu} \mathcal{D}_\nu \Psi\right)-\mu^2\Psi=0,\quad \frac{1}{\sqrt{-g}} \mathcal{D}^*_\mu \left(\sqrt{-g} g^{\mu\nu} \mathcal{D}^*_\nu \Psi^*\right)-\mu^2\Psi^*=0.
\end{align}
Since the transformed metric in Eq.\,(\ref{eq:transfmetrc01}) is stationary with translation symmetries on coordinates $T$ and $\Phi$, the ansatz to Eq.\,(\ref{eq:fieldequation01}) is
\begin{align}
\Psi(T,r,\theta,\Phi)=e^{-i\omega T}e^{im\Phi} R(r) \Theta(\theta),
\end{align}
where $\omega$ and $m$ are the frequency and angular number, respectively. Then, Eq.\,(\ref{eq:fieldequation01}) is separable into radial and $\theta$-directional equations with a separate variable $\mathcal{K}$. The radial equation is
\begin{align}\label{eq:radialeq01}
\frac{1}{R(r)}\partial_r \left(\Delta_r \partial_r R(r) \right)+\frac{1}{\Delta_r}\left(\omega(r^2+a^2) -am \left(1-\frac{1}{3}\Lambda r^2\right)-qQr\right)^2-\mu^2r^2-\mathcal{K}=0.
\end{align}
Furthermore, the $\theta$-directional equation is found in the generalized scalar hyper-spheroidal equation for a massive scalar field\cite{Berti:2005gp,Cho:2009wf}.
\begin{align}\label{eq:thetadirectionaleq}
\frac{1}{\sin\theta \,\Theta(\theta)}\partial_\theta \left(\sin\theta \Delta_\theta \partial_\theta \Theta(\theta) \right)-\frac{1}{\Delta_\theta}\left(a\omega\sin\theta-m\Delta_\theta \csc\theta\right)^2-a^2 \mu^2\cos^2\theta+\mathcal{K}=0,
\end{align}
which can also be rewritten in a known form called the Heun's equation\cite{Suzuki:1998vy,Suzuki:1999nn,Yoshida:2010zzb,Kraniotis:2016maw}.

Fortunately, if we focus on the scalar field fluxes at the spatial boundaries, we require only the normalized condition of the solution to Eq.\,(\ref{eq:thetadirectionaleq}) rather than its various properties. The normalized condition is\cite{Cho:2009wf}
\begin{align}
\int \Theta(\theta)\Theta^*(\theta) d\Omega_2=1.
\end{align}
When the solution is integrated to find the fluxes, it simply yields unity. Note that the generalized scalar hyper-spheroidal equation becomes a scalar spheroidal equation with an angular momentum number $\mathcal{K}=\ell (\ell+1)$ for Kerr and Schwarzschild-de Sitter black holes\cite{Teukolsky:1973ha,Berti:2005gp,Pappas:2016ovo}. The case of the KN(A)dS black holes corresponds to the numerical solution to the Heun's equation discussed in \cite{Suzuki:1998vy,Suzuki:1999nn,Yoshida:2010zzb,Kraniotis:2016maw}. Moreover, the separate variable $\mathcal{K}$ becomes a non-integer number close to $\ell (\ell+1)$\cite{Zhang:2014kna,Delice:2015zga}.  

We now consider our main focus, the radial equation in Eq.\,(\ref{eq:radialeq01}). Note that the radial equation is only solvable when the boundary conditions are determined at both spatial ends. Hence, both spatial boundaries are physically connected. Considering this connection, we investigate the relationship between changes in the black hole and the boundary conditions. First, we study one of the boundaries at the outer horizon, the radial solution, because it is commonly applied to any case of cosmological constant. Note that we discuss another boundary in the following sections. The tortoise coordinate is defined to solve Eq.\,(\ref{eq:radialeq01}), where
\begin{align}\label{eq:tortoisecoordinate}
\frac{dr^*}{dr}=\frac{r^2+a^2}{\Delta_r}.
\end{align}
The interval of the tortoise coordinate depends on the sign of the cosmological constant. For the asymptotically flat case, $\Lambda=0$, the radial interval $(r_\text{h},+\infty)$ becomes $(-\infty,+\infty)$ in the tortoise coordinate, while for the asymptotically AdS case, $\Lambda<0$, the radial interval $(r_\text{h},+\infty)$ becomes $(-\infty,0)$ in the tortoise coordinate. For the dS case, $\Lambda>0$, the radial interval $(r_\text{h},r_\text{c})$ becomes $(-\infty,+\infty)$ in the tortoise coordinate. Using Eq.\,(\ref{eq:tortoisecoordinate}), we rewrite the radial equation in Eq.\,(\ref{eq:radialeq01}) in terms of the tortoise coordinate, such that
\begin{align}\label{eq:radialeqtortoise01}
&\frac{1}{R(r^*)}\frac{d^2 R(r^*)}{d{r^*}^2}+\frac{2r\Delta_r }{(r^2+a^2)^2R(r^*)}\frac{d R(r^*)}{dr^*}+\left(\omega -\frac{a m\left(1-\frac{1}{3}\Lambda r^2\right)}{(r^2+a^2)}-\frac{qQr}{(r^2+a^2)}\right)^2-\frac{(\mu^2r^2+\mathcal{K})\Delta_r }{(r^2+a^2)^2}=0,
\end{align}
which can be expressed as a Schr\"{o}dinger-type equation via a transformation.
\begin{align}
R(r^*)\rightarrow \frac{\mathcal{R}(r^*)}{\sqrt{r^2+a^2}}.
\end{align}
Then, the radial equation in Eq.\,(\ref{eq:radialeqtortoise01}) becomes
\begin{align}\label{eq:radialeqtortoise03}
&\frac{1}{\mathcal{R}(r^*)}\frac{d^2 \mathcal{R}(r^*)}{d{r^*}^2}+\left(\omega-\frac{am \left(1-\frac{1}{3}\Lambda r^2\right)}{r^2+a^2} -\frac{qQr}{r^2+a^2}\right)^2-\frac{\Delta_r (r d\Delta_{r}-2\Delta_r)}{(r^2+a^2)^3}-\frac{3a^2\Delta_r^2}{(r^2+a^2)^4}\\
&-\frac{(\mu^2r^2+\mathcal{K})\Delta_r }{(r^2+a^2)^2}=0,\quad d\Delta_{r} \equiv \frac{\partial \Delta_r}{\partial r}=-2M-\frac{2}{3}r(r^2+a^2)\Lambda+2r\left(1-\frac{1}{3}\Lambda r^2\right).\nonumber
\end{align}
Flowing into the outer horizon, the conserved scalar field quantities cannot be measured separately from those of the black hole. Hence, the conserved scalar field quantities flowing into the black hole must be assumed to be those of the black hole. The magnitudes of the conserved quantities are determined by the scalar field fluxes at the outer horizon, and the fluxes can be obtained from the scalar field solution at the outer horizon. At the outer horizon, the radial equation of Eq.\,(\ref{eq:radialeqtortoise03}) becomes
\begin{align}
\frac{1}{\mathcal{R}(r^*)}\frac{d^2 \mathcal{R}(r^*)}{d{r^*}^2}+\left(\omega -m\Omega_\text{h}-q\Phi_\text{h}\right)^2=0.
\end{align}
The radial solution is simply
\begin{align}
\mathcal{R}(r^*)\sim e^{\pm i (\omega-m\Omega_\text{h}-q\Phi_\text{h})r^*},\quad r^*\rightarrow -\infty\,\,(r\rightarrow r_\text{h}),
\end{align}
where the positive and negative signs correspond to the outgoing and ingoing waves, respectively. Classically, the outer horizon of a black hole is a one-way surface in which there are no outgoing null or time-like geodesics. Thus, we select a negative sign for the ingoing wave of this classical scalar field. Furthermore, the outgoing wave is not smooth at the outer horizon and, thus, the regularity condition ensures a purely ingoing wave\cite{Berti:2009kk,Brito:2015oca}. Then, the radial solution can be obtained as
\begin{align}\label{eq:radialsolution05}
\mathcal{R}(r^*)=\mathcal{T}e^{- i (\omega-m\Omega_\text{h}-q\Phi_\text{h})r^*},
\end{align}
where $\mathcal{T}$ is the transmission amplitude. Therefore, the scalar field solution and its conjugate at the event horizon are
\begin{align}\label{eq:fluxes01}
&\Psi(T,r^*,\theta,\Phi)=\frac{\mathcal{T}}{\sqrt{r^2_\text{h}+a^2}} e^{-i\omega T} e^{-i(\omega-m\Omega_\text{h}-q\Phi_\text{h})r^*} \Theta(\theta) e^{im\Phi},\\
&\Psi^*(T,r^*,\theta,\Phi)=\frac{\mathcal{T^*}}{\sqrt{r^2_\text{h}+a^2}} e^{i\omega T} e^{i(\omega-m\Omega_\text{h}-q\Phi_\text{h})r^*} \Theta^*(\theta) e^{-im\Phi}.\nonumber
\end{align}
As these solutions are obtained from the field equation limited at the outer horizon, they constitute exact solutions at the outer horizon only. Despite this limitation, however, these solutions are sufficient for our analysis of the conserved quantities of the scalar field flowing into the black hole.

\section{Scalar Field Fluxes}\label{sec4}

The conserved quantities of the scalar field, such as energy, angular momentum, and electric charge, flow into the black hole. Since the scalar field configuration cannot be measured inside the outer horizon, when flowing through the outer horizon, the conserved quantities of the scalar field can be assumed to be merged with and, hence, measured as those of the black hole. Then, the conserved quantities transferred into the black hole can be obtained from the scalar field fluxes at the outer horizon expressed in Eq.\,(\ref{eq:fluxes01}). The scalar field fluxes are obtained by integrating the energy-momentum tensor.
\begin{align}
T^\mu_\nu=\frac{1}{2}\mathcal{D}^\mu\Psi \partial_\nu \Psi^*+\frac{1}{2}\mathcal{D}^{*\mu}\Psi^* \partial_\nu \Psi-\delta^\mu_\nu\left(\frac{1}{2}\mathcal{D}_\mu\Psi \mathcal{D}^{*\mu}\Psi^* +\frac{1}{2}\mu^2\Psi\Psi^*\right).
\end{align}
Subsequently, at the outer horizon, the energy, angular momentum, and electric charge fluxes are, respectively,
\begin{align}
\frac{dE}{dT}&=\frac{4\pi|\mathcal{T}|^2}{\Xi}\omega(\omega-m\Omega_\text{h}-q\Phi_\text{h}),\\
\frac{dL}{dT}&=\frac{4\pi|\mathcal{T}|^2}{\Xi}m(\omega-m\Omega_\text{h}-q\Phi_\text{h}),\nonumber\\
\frac{dq}{dT}&=\frac{4\pi|\mathcal{T}|^2}{\Xi}q(\omega-m\Omega_\text{h}-q\Phi_\text{h}).\nonumber
\end{align}
These indicate that the magnitudes of the scalar field fluxes are also associated with the absolute transmission amplitude. The fluxes at the outer horizon indicate that the conserved quantities of the scalar field flow into the black hole and thus they cannot be measured by an outside observer. Instead, the observer can find their contributions to the black hole by measuring the mass, angular momentum, and electric charge. Then, when we consider an infinitesimal time interval $dT$, the changes in the black hole can be obtained from the fluxes, such that 
\begin{align}\label{eq:changeblackhole01}
dM_\text{B}&=\frac{4\pi|\mathcal{T}|^2}{\Xi}\omega(\omega-m\Omega_\text{h}-q\Phi_\text{h})dT,\\
dJ_\text{B}&=\frac{4\pi|\mathcal{T}|^2}{\Xi}m(\omega-m\Omega_\text{h}-q\Phi_\text{h})dT,\nonumber\\
dQ_\text{B}&=\frac{4\pi|\mathcal{T}|^2}{\Xi}q(\omega-m\Omega_\text{h}-q\Phi_\text{h})dT.\nonumber
\end{align}
The state of the KN(A)dS black hole is determined by the mass, angular momentum, and electric charge. As the scalar field changes these three variables according to Eq.\,(\ref{eq:changeblackhole01}), we should consider the effects of the changes in the three conserved quantities to determine the final state of the black hole. The changes in the black hole are herein considered infinitesimal. Therefore, we consider the series expansion of the changes and focus on the first order. In fact, compared with the magnitudes of black hole quantities, scalar field effects are infinitesimally small. Hence, rather than the nonlinear terms, the first-order term is the most dominant term for estimating the changes in the black hole due to scalar field scattering. Moreover, when considering the changes during the infinitesimal time interval, we focus on the case in which the first-order term is significant.

\section{Weak Cosmic Censorship Conjecture in Near-Extremal KN(A)dS Black Holes}\label{sec5}

We herein investigate the validity of the WCC conjecture for near-extremal KN(A)dS black holes. Our analysis also includes the {\it{extremal}} cases for {\it{arbitrary}} cosmological constant. The sign of the cosmological constant determines the asymptotic geometry, which is a significant reason for choosing asymptotic boundary conditions. First, we study KN black holes with $\Lambda=0$, i.e., the asymptotically flat case. Then, we move to KNAdS black holes with $\Lambda<0$ and KNdS black holes with $\Lambda>0$. The near-Nariai black hole in the KNdS case is also considered.

The WCC conjecture predicts that no naked singularity can be observed by a static observer. In the spacetime of a KN(A)dS black hole, the naked singularity is only observed when the inner and outer horizons disappear owing to an overcharged black hole. Note that we refer to overcharging, including overspinning. Hence, we focus on whether the KN(A)dS black hole can be {\it{overcharged}} by a massive scalar field with an electric charge. The scalar field carries its conserved quantities into the black hole. During the infinitesimal time interval, the carried, conserved quantities change the black hole properties according to the fluxes given in Eq.\,(\ref{eq:changeblackhole01}). In other words, the state of the black hole in terms of $(M_\text{B},J_\text{B},Q_\text{B})$ varies as $(M_\text{B}+dM_\text{B},J_\text{B}+dJ_\text{B},Q_\text{B}+dQ_\text{B})$. As the locations of the inner and outer horizons are determined by the function $\Delta_r$ of the three variables $(M_\text{B},J_\text{B},Q_\text{B})$, we can determine whether the black hole becomes a naked singularity by observing the change in $\Delta_r$ under the variation due to the scalar field. If a horizon remains in the changed $\Delta_r$, the WCC is valid. Otherwise, it is violated. Recall that our analysis is based on the changes caused by the scattering of the scalar field, which has minimal energy compared with the black hole. Hence, the change in the black hole is also infinitesimal. Therefore, we study near-extremal black holes that can potentially be overcharged by the small variations resulting from the scalar field.

Detailed investigations of the WCC conjecture depend on the asymptotic geometries, which are determined by the sign of the cosmological constant.

\subsection{Near-Extremal KN Black Holes: $\Lambda=0$}\label{sec:sec52a}

\begin{figure}[h]
\centering
\subfigure[{$\Delta_r$ with $(M_\text{B},J_\text{B},Q_\text{B})$.}] {\includegraphics[scale=0.8,keepaspectratio]{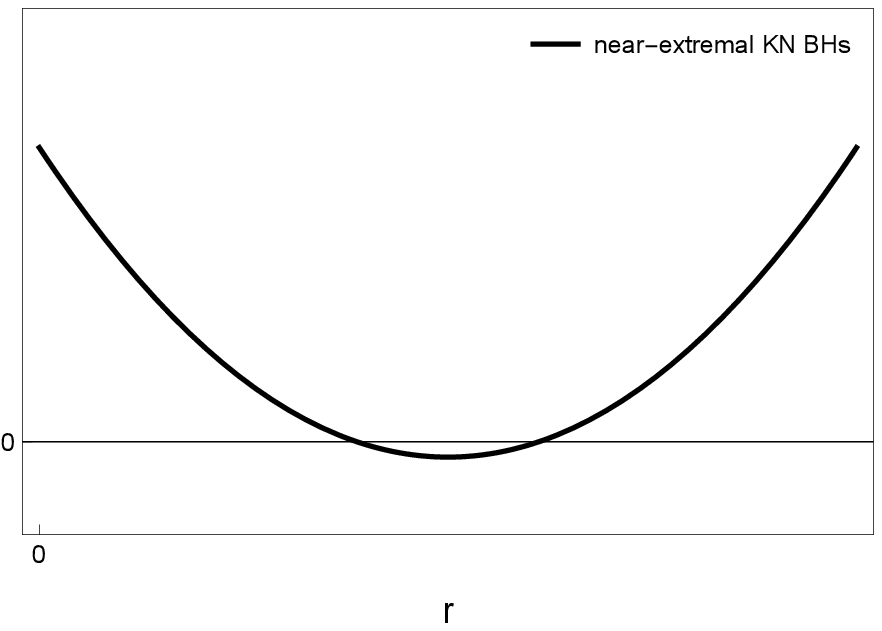}}\quad
\subfigure[{$\Delta_r$ with $(M_\text{B}+dM_\text{B},J_\text{B}+dJ_\text{B},Q_\text{B}+dQ_\text{B})$.}] {\includegraphics[scale=0.8,keepaspectratio]{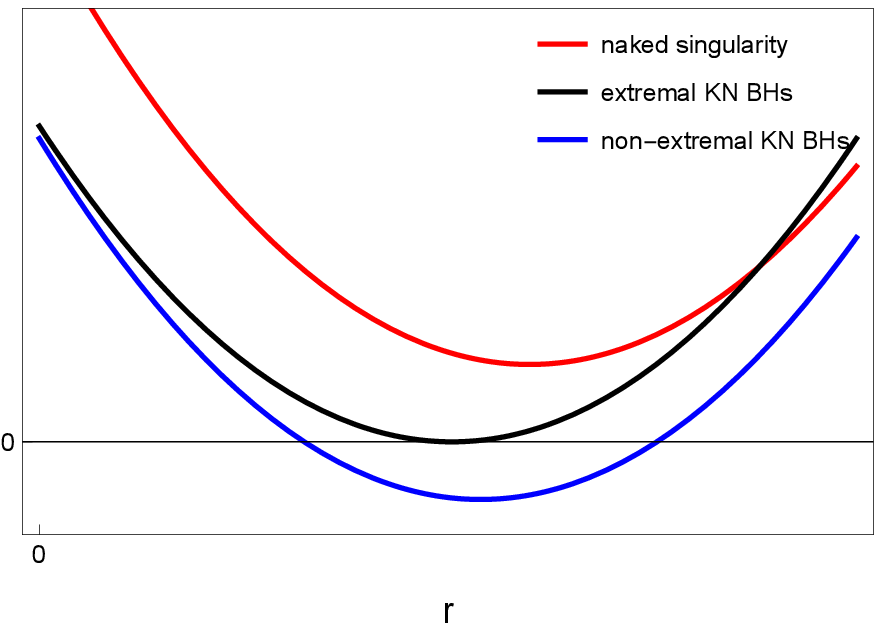}}
\caption{{\small $\Delta_r$ with $\Lambda=0$ in the initial and final states.}}
\label{fig:number1}
\end{figure}

As previously explained, the WCC conjecture can be tested by assessing the change in $\Delta_r$. As the near-extremal black hole is assumed to be in the initial state, $\Delta_r$ with $(M_\text{B},J_\text{B},Q_\text{B})$ is the same as in Fig.\,\ref{fig:number1}\,(a).
The function $\Delta_r$ has a minimum with a small negative value $\Delta_\text{min}$ owing to the near-extremal condition. The solutions to $\Delta_r=0$ determine the locations of the inner and outer horizons. Hence, two horizons are located near the minimum. The conserved quantities carried by the scalar field change the function $\Delta_r$ according to $(M_\text{B}+dM_\text{B},J_\text{B}+dJ_\text{B},Q_\text{B}+dQ_\text{B})$ during the infinitesimal time interval in the final state, as shown in Fig.\,\ref{fig:number1}\,(b). The possible final states of the black hole can be categorized into two cases based on the minimum value of the function $\Delta_r$: a black hole or naked singularity. If the minimum value is negative or zero, $\Delta_r$ always has at least one solution. Thus, the singularity cannot be observed, owing to the horizon. Additionally, a positive minimum is a naked singularity with no horizon. Thus, we can determine the validity of the WCC conjecture based on the {\it{sign}} of the minimum value in the final state.

In the initial state, the location of the minimum of the function $\Delta_r$ is denoted as $r_\text{min}$. Then, the initial state with the near-extremal condition is
\begin{align}\label{eq:kerrinitial01}
\Delta_\text{min}\equiv \left.\Delta_r\right|_{r=r_\text{min}},\quad -\Delta_\text{min}&\ll 1,\quad \left.\frac{\partial \Delta_r}{\partial r}\right|_{r=r_\text{min}}=\frac{\partial \Delta_\text{min}}{\partial r_\text{min}}=0,\quad \left.\frac{\partial^2 \Delta_r}{\partial r^2}\right|_{r=r_\text{min}}=\frac{\partial^2 \Delta_\text{min}}{\partial r_\text{min}^2}>0,
\end{align}
where the black hole is in the $(M_\text{B},J_\text{B},Q_\text{B})$ state. Then, during the time interval $dT$, the conserved quantities are carried to the black hole by the scattered scalar field. The variations in the black hole are represented as $(dM_\text{B},dJ_\text{B},dQ_\text{B})$, which imply that the final state of the black hole is $(M_\text{B}+dM_\text{B},J_\text{B}+dJ_\text{B},Q_\text{B}+dQ_\text{B})$. As this change also alters the function $\Delta_r$, the value of the minimum can differ from its initial value. For the near-extremal case given in Eq.\,(\ref{eq:kerrinitial01}), the change in the minimum value is obtained as the first order in terms of the infinitesimal variations in the mass, angular momentum, and electric charge.
\begin{align}\label{eq:WCCanalysis05}
d\Delta_\text{min}&\equiv \Delta_\text{min}(M_\text{B}+dM_\text{B},J_\text{B}+dJ_\text{B},Q_\text{B}+dQ_\text{B},r_\text{min}+dr_\text{min})-\Delta_\text{min}(M_\text{B},J_\text{B},Q_\text{B},r_\text{min})\\
&=\frac{\partial \Delta_\text{min}}{\partial  M_\text{B}}dM_B+\frac{\partial \Delta_\text{min}}{\partial  J_\text{B}}dJ_\text{B}+\frac{\partial \Delta_\text{min}}{\partial  Q_\text{B}}dQ_\text{B}+\frac{\partial\Delta_\text{min}}{\partial  r_\text{min}}dr_\text{min}.\nonumber
\end{align}
The location of the minimum is also determined by the independent variables $(M_\text{B},J_\text{B},Q_\text{B})$. Thus, $r_\text{min}$ also moves into $r_\text{min}+dr_\text{min}$. The change in location is rewritten in terms of $(dM_\text{B},dJ_\text{B},dQ_\text{B})$. Here, we consider the near-extremal black hole for which the minimum value is nonzero. Then, the minimum value of the final state is the sum of the initial value and its change. Based on Eq.\,(\ref{eq:kerrinitial01}),
\begin{align}
\Delta_\text{min}+d\Delta_\text{min}&=\Delta_\text{min}+4\pi|\mathcal{T}|^2(\omega-m\Omega_\text{h}-q\Phi_\text{h})\left(\omega+m\Omega_\text{eff}+q\Phi_\text{eff}\right)\left(\frac{\partial \Delta_\text{min}}{\partial M_\text{B}}\right)dT,
\end{align}
where
\begin{align}
\Omega_\text{eff}\equiv\left.{\frac{\partial \Delta_\text{min}}{\partial J_\text{B}}}\right/{\frac{\partial \Delta_\text{min}}{\partial M_\text{B}}},\quad \Phi_\text{eff}\equiv \left.{\frac{\partial \Delta_\text{min}}{\partial J_\text{B}}}\right/{\frac{\partial \Delta_\text{min}}{\partial M_\text{B}}}.
\end{align}
To rewrite the location of the minimum $r_\text{min}$ in terms of the outer horizon $r_\text{h}$, we assume that the locations of the minimum and horizon are very close, and we denote their small displacement by $\epsilon$. Then, the initial minimum value can be rewritten in terms of $r_\text{h}$ and $\epsilon$.
\begin{align}\label{eq:epsilonandminimum}
r_\text{h}-r_\text{min}=\epsilon\ll 1,\quad \Delta_\text{min}=-\epsilon^2. 
\end{align}
According to Eq.\,(\ref{eq:epsilonandminimum}),
\begin{align}\label{eq:kerrinitial03}
&\Omega_\text{eff}=-\Omega_\text{h}-\frac{2 a r_\text{h}}{(r_\text{h}^2+a^2)^2}\epsilon+\mathcal{O}(\epsilon^2),\quad \Phi_\text{eff}=-\Phi_\text{h}-\frac{Q(r_\text{h}^2-a^2)}{(r_\text{h}^2+a^2)^2}\epsilon+\mathcal{O}(\epsilon^2),\\
&{\frac{\partial \Delta_\text{min}}{\partial M_\text{B}}}=-\frac{2(r_\text{h}^2+a^2)}{r_\text{h}}+2 \left(1-\frac{a^2}{r_\text{h}^2}\right)\epsilon+\mathcal{O}(\epsilon^2).\nonumber
\end{align}
Using Eqs.\,(\ref{eq:kerrinitial01}) and (\ref{eq:kerrinitial03}), the final minimum value for the near-extremal black hole can be obtained as
\begin{align}\label{eq:kerrwccminimum}
&\Delta_\text{min}+d\Delta_\text{min}=-\frac{8\pi|\mathcal{T}|^2(r_\text{h}^2+a^2) (\omega-m\Omega_\text{h}-q\Phi_\text{h})^2}{r_\text{h}}dT + \mathcal{O}(\epsilon).
\end{align}
As we consider the changes in the infinitesimal time interval, $dT\ll 1$, the first order of $\epsilon$ is negligible in Eq.\,(\ref{eq:kerrwccminimum}), because every term includes $dT$. Hence, the first term in Eq.\,(\ref{eq:kerrwccminimum}) is dominant under this scattering process. This outcome indicates that the minimum value is still negative in the final state. Moreover, the negativity is {\it{independent}} of the scalar field configuration determined by $(\mathcal{T},\omega, m, q)$. In particular, the absolute value of the transmission amplitude implies that the negativity is also independent of the asymptotic boundary condition. Therefore, the WCC conjecture is valid for near-extremal KN black holes, including the extremal case. For the extremal black hole, the final minimum value is exactly the first term in Eq.\,(\ref{eq:kerrwccminimum}). Hence, the WCC conjecture is {\it{valid}} owing to the negative minimum value.

The validity of the WCC conjecture in Eq.\,(\ref{eq:kerrwccminimum}) is an independent consequence of the scalar field modes. However, the transmission amplitude is related to the other amplitudes at the asymptotic boundary. To determine the relationship between the amplitudes, the asymptotic solutions should be considered. For the spatially asymptotic region with $\Lambda=0$, the Schr\"{o}dinger-type equation in Eq.\,(\ref{eq:radialeqtortoise03}) is rewritten as
\begin{align}
\frac{1}{\mathcal{R}(r^*)}\frac{d R(r^*)}{dr^*}+(\omega^2-\mu^2)=0.
\end{align}
The asymptotic solutions are simply obtained as
\begin{align}
\mathcal{R}(r^*)\sim e^{\pm i \sqrt{\omega^2-\mu^2}r^*},\quad r^*\rightarrow +\infty\,\,(r\rightarrow +\infty),
\end{align}
where we consider $\omega>\mu$ and $\mu\geq 0$ to maintain a real value for the square root term. Then, the asymptotic radial solution is obtained as
\begin{align}
\mathcal{R}(r^*)=\mathcal{O}e^{i \sqrt{\omega^2-\mu^2}r^*}+\mathcal{I}e^{-i \sqrt{\omega^2-\mu^2}r^*},
\end{align}
where $\mathcal{O}$ and $\mathcal{I}$ are the amplitudes of the outgoing and incident waves at the spatial infinity, respectively. As the solution and its conjugate in Eq.\,(\ref{eq:radialeqtortoise03}) are linearly independent, their Wronskian produces an independent value under the tortoise coordinate, such that
\begin{align}
W[\mathcal{R}(r^*),\mathcal{R}^*(r^*)]=\frac{d \mathcal{R}}{d r^*} \mathcal{R}^* - \mathcal{R} \frac{d \mathcal{R}^*}{d r^*}.
\end{align}
The Wronskian should be coincident at both boundaries.
\begin{align}
\lim_{r^*\rightarrow -\infty} W[\mathcal{R}(r^*),\mathcal{R}^*(r^*)]=\lim_{r^*\rightarrow +\infty} W[\mathcal{R}(r^*),\mathcal{R}^*(r^*)].
\end{align}
Hence, we find that
\begin{align}\label{eq:fluxanalysis01}
|\mathcal{O}|^2=|\mathcal{I}|^2-\frac{\omega-m\Omega_\text{h}-q\Phi_\text{h}}{\sqrt{\omega^2-\mu^2}}|\mathcal{T}|^2.
\end{align}
This shows that the net flux of the scattering is influx, $|\mathcal{O}|^2<|\mathcal{I}|^2$, for
\begin{align}\label{eq:eq:fluxanalysis02}
\omega>\mu\,\,\text{and}\,\,\omega>m\Omega_\text{h}+q\Phi_\text{h},
\end{align}
and that the net flux is superradiant, $|\mathcal{O}|^2>|\mathcal{I}|^2$, for
\begin{align}\label{eq:eq:fluxanalysis03}
\mu<\omega<m\Omega_\text{h}+q\Phi_\text{h}.
\end{align}
Combining Eqs.\,(\ref{eq:fluxanalysis01}), (\ref{eq:eq:fluxanalysis02}), and (\ref{eq:eq:fluxanalysis03}), the transmission per unit net amplitude is obtained as
\begin{align}\label{eq:ratiofluxes05}
\frac{|\mathcal{T}|^2}{|\mathcal{F}|^2}\equiv \frac{|\mathcal{T}|^2}{||\mathcal{I}|^2-|\mathcal{O}|^2|}=\pm\frac{\sqrt{\omega^2-\mu^2}}{\omega-m\Omega_\text{h}-q\Phi_\text{h}},
\end{align}
where $|\mathcal{F}|^2\equiv ||\mathcal{I}|^2-|\mathcal{O}|^2|$. This net amplitude represents the magnitude of the net ingoing or outgoing flux measured by the asymptotic observer. For Eq.\,(\ref{eq:ratiofluxes05}) to positive, the positive and negative signs must correspond to absorption scattering and superradiance, respectively.

Here, we focus on the change in the black hole owing to a unit net flux measured at the spatial infinity. Hence, the amplitudes are normalized to set the magnitude of the net flux $|\mathcal{F}|^2$ to unity rather than the detailed form of the amplitudes. The normalized amplitudes are
\begin{align}\label{eq:netfluxtransfm01}
\frac{|\mathcal{T}|^2}{|\mathcal{F}|^2}\rightarrow |\hat{\mathcal{T}}|^2,\quad \frac{|\mathcal{O}|^2}{|\mathcal{F}|^2}\rightarrow |\hat{\mathcal{O}}|^2,\quad \frac{|\mathcal{I}|^2}{|\mathcal{F}|^2}\rightarrow |\hat{\mathcal{I}}|^2.
\end{align}
These transformations show the changes in the black hole per unit magnitude of the net flux. Based on Eq.\,(\ref{eq:netfluxtransfm01}), the change in the minimum value of Eq.\,(\ref{eq:kerrwccminimum}) is rewritten in terms of the unit magnitude of the net flux measured at the asymptotic infinity.
\begin{align}\label{eq:netfluxtransfm03}
\Delta_\text{min}+d\Delta_\text{min}&\approx-\frac{8\pi|\hat{\mathcal{T}}|^2(r_\text{h}^2+a^2) (\omega-m\Omega_\text{h}-q\Phi_\text{h})^2}{r_\text{h}}dT\nonumber\\
&=\mp \frac{8\pi(r_\text{h}^2+a^2)\sqrt{\omega^2-\mu^2} (\omega-m\Omega_\text{h}-q\Phi_\text{h})}{r_\text{h}}dT,
\end{align}
where negative and positive signs correspond to absorption and superradiance, respectively. In accordance with Eq.\,(\ref{eq:netfluxtransfm03}), the change in the minimum is associated with $\sqrt{\omega^2-\mu^2}(\omega-m\Omega_\text{h}-q\Phi_\text{h})$ for a given near-extremal black hole and is negative for any scalar field modes. Thus, the WCC conjecture is valid under {\it{any boundary condition}} of the asymptotic region.

\subsection{Near-Extremal KN-AdS and KN-dS Black Holes: $\Lambda\not=0$}

Here, we generalize our analysis presented in Sec.\,\ref{sec:sec52a} to black holes with nonzero cosmological constants. In these cases, the asymptotic geometries differ from those of the KN black holes and play a significant role in determining the boundary conditions. For negative and positive cosmological constants, the spacetimes are asymptotically the AdS and dS boundaries, respectively. Because the boundary condition at the asymptotic region is associated with the transmission amplitude at the outer horizon, we can determine the validity of the WCC conjecture and its association with the boundary conditions of both KN-AdS and KN-dS black holes through analysis of the outer horizon. We also discuss detailed changes in near-extremal KN-dS black holes regarding the Kerr-type and Nariai-type extremal cases.

\subsubsection{Kerr-Type Near-Extremal KN-AdS and KN-dS Black Holes}

\begin{figure}[h]
\centering
\subfigure[{$\Delta_r$ with $\Lambda<0$.}] {\includegraphics[scale=0.8,keepaspectratio]{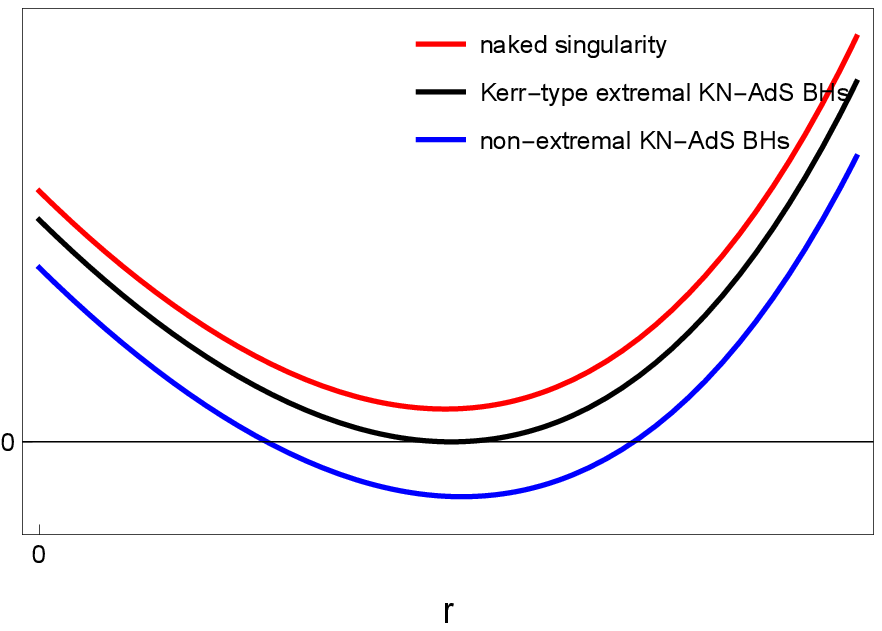}}\quad
\subfigure[{$\Delta_r$ with $\Lambda>0$.}] {\includegraphics[scale=0.8,keepaspectratio]{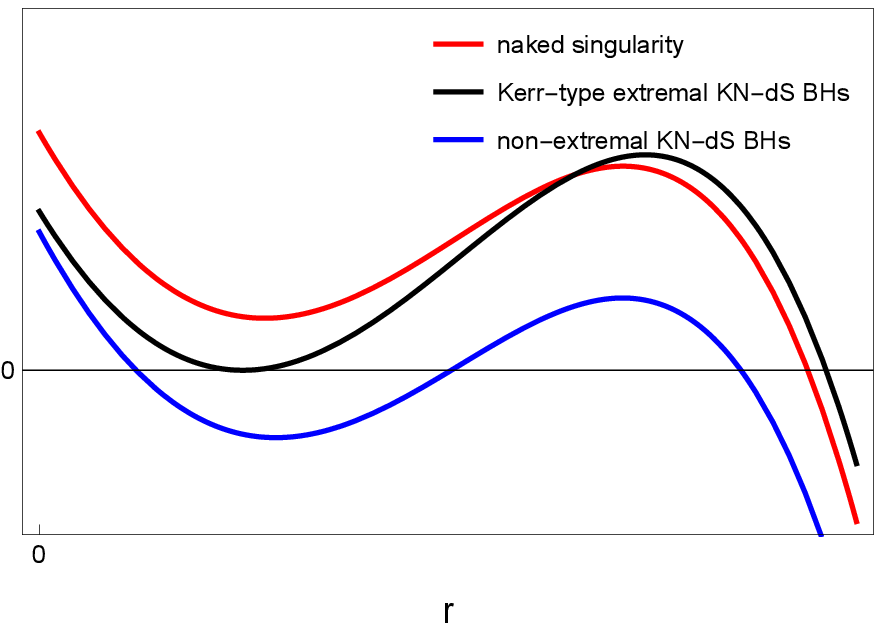}}
\caption{{\small The states of KN-(A)dS black holes by $\Delta_r$ with $\Lambda\not=0$.}}
\label{fig:number2}
\end{figure}

In the Kerr-type extremal case, the inner and outer horizons coincide. This type of extremal black hole is commonly observed for negative and positive cosmological constants, as shown in Fig.\,\ref{fig:number2}.
Hence, we maintain the general value of the cosmological constant for the Kerr-type near-extremal cases. The analysis procedure is analogous to that given in between Eqs.\,(\ref{eq:WCCanalysis05}) and (\ref{eq:kerrwccminimum}). The black hole with the Kerr-type extremal condition has one horizon at the minimum of the function $\Delta_r$. Then, the initial state of the Kerr-type near-extremal black hole is identical to that expressed in Eq.\,(\ref{eq:WCCanalysis05}). The scalar field scattering carries the conserved quantities that change the black hole state according to the fluxes in Eq.\,(\ref{eq:changeblackhole01}). The change in the minimum value is
\begin{align}\label{eq:Kerrminimum10}
\Delta_\text{min}+d\Delta_\text{min}=\Delta_\text{min}+\frac{4\pi}{\Xi}(\omega-m\Omega_\text{h}-q\Phi_\text{h})(\omega-m\Omega_\text{eff}-q\Phi_\text{eff})|\mathcal{T}|^2 \frac{\partial \Delta_\text{min}}{\partial M_\text{B}}dT.
\end{align}
Under the near-extremal condition, $r_\text{h}-r_\text{min}=\epsilon\ll 1$, and we have
\begin{align}\label{eq:Kerrminimum12}
\Omega_\text{eff}&=-\Omega_\text{h}-\frac{2 a r_\text{h} \Xi}{(r_\text{h}^2+a^2)^2}\epsilon+\mathcal{O}(\epsilon^2),\quad \Phi_\text{eff}=-\Phi_\text{h}-\frac{Q(r_\text{h}^2-a^2)}{(r_\text{h}^2+a^2)^2}\epsilon+\mathcal{O}(\epsilon^2),\\
{\frac{\partial \Delta_\text{min}}{\partial M_\text{B}}}&=-\frac{2(r_\text{h}^2+a^2)\Xi}{r_\text{h}}+2\Xi \left(1-\frac{a^2}{r_\text{h}^2}\right)\epsilon+\mathcal{O}(\epsilon^2).\nonumber
\end{align}
The initial minimum is also rewritten to the second order of $\epsilon$.
\begin{align}\label{eq:Kerrminimum15}
\Delta_\text{min}=\frac{1}{3}(-3+a^2\Lambda + 6r_\text{h}^2\Lambda)\epsilon^2-\frac{8}{3}r_\text{h}\Lambda\epsilon^3+\Lambda \epsilon^4 + \mathcal{O}(\epsilon^5). 
\end{align}
Then, using Eqs.\,(\ref{eq:Kerrminimum10}), (\ref{eq:Kerrminimum12}), and (\ref{eq:Kerrminimum15}), the minimum value of the final state is obtained as
\begin{align}\label{eq:Kerrminimum20}
\Delta_\text{min}+d\Delta_\text{min}=&-\frac{8\pi|\mathcal{T}|^2(r_\text{h}^2+a^2) (\omega-m\Omega_\text{h}-q\Phi_\text{h})^2}{r_\text{h}}dT\\
&+\frac{8  \pi |\mathcal{T}|^2 (\omega-m\Omega_\text{h}-q\Phi_\text{h})}{r_\text{h}^2(r_\text{h}^2+a^2)}(a^2 q Q r_\text{h}+2a m r_\text{h}^2 \Xi -a^4 (\omega- m \Omega_\text{h} -q \Phi_\text{h})\nonumber\\
&-r_\text{h}^3 (q(Q+r_\text{h}\Phi_\text{h})-r_\text{h}\omega + m r_\text{h} \Omega_\text{h}))\epsilon dT + \mathcal{O}(\epsilon^2).\nonumber
\end{align}
Since we consider an infinitesimal time interval $dT$, the first order of $\epsilon$ is negligible. Therefore, according to the leading term in Eqs.\,(\ref{eq:Kerrminimum20}), the change in the minimum value is {\it{negative}}. This implies that the Kerr-type near-extremal KN-AdS and KN-dS black holes cannot be overcharged by the scalar field scattering. Thus, the WCC conjecture is {\it{valid}}.  

Furthermore, the transmission amplitude is the {\it{only}} term associated with the asymptotic boundary condition in Eq.\,(\ref{eq:Kerrminimum20}). To determine the transmission amplitude, we must impose a boundary condition at the outer horizon and the asymptotic boundary, and solve the radial equation in Eq.\,(\ref{eq:radialeqtortoise03}). However, the detailed value of the transmission amplitude does not change the negativity of the final minimum value, because the transmission amplitude is given by Eq.\,(\ref{eq:Kerrminimum20}) as {\it{the square of the absolute value}}. Hence, the negativity of the final minimum value is preserved for any transmission amplitude. This implies that an asymptotic boundary condition does not change the negativity. Therefore, the WCC conjecture is valid for {\it{any mode and any boundary condition}} of the scalar field for {\it{both positive and negative}} cosmological constants.

\subsubsection{Nariai-Type Near-Extremal KN-dS Black Holes}

We herein investigate the change in the state of a Nariai-type near-extremal KN-dS black hole. The Nariai-type extremal condition occurs when the outer and cosmological horizons are coincident. This case has a weaker relationship with the WCC conjecture but exhibits some interesting behaviors under scattering.

\begin{figure}[h]
\centering
\subfigure[{$\Delta_r$ in the initial state.}] {\includegraphics[scale=0.8,keepaspectratio]{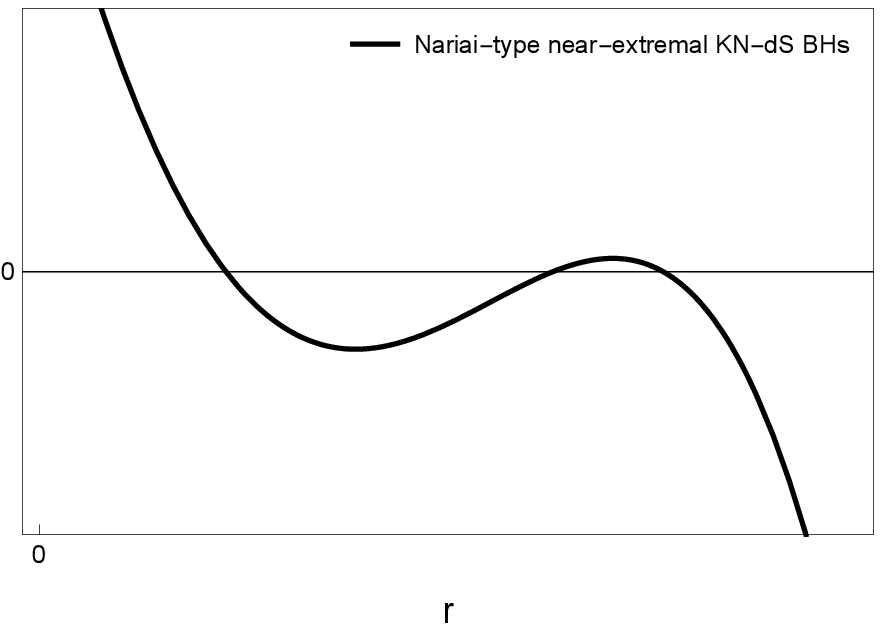}}\quad
\subfigure[{$\Delta_r$ in possible final states.}] {\includegraphics[scale=0.8,keepaspectratio]{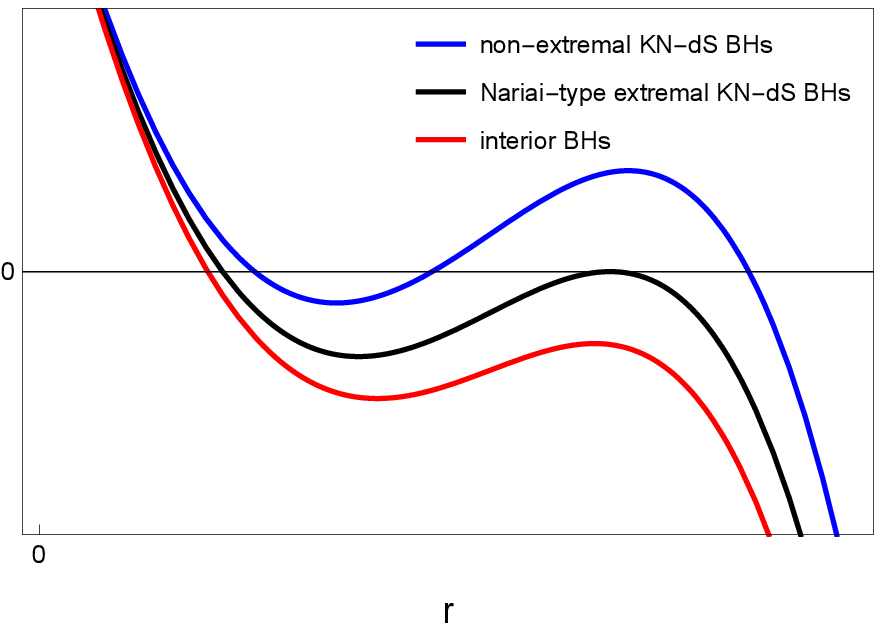}}
\caption{{\small The states of KN-dS black holes by $\Delta_r$ with $\Lambda>0$.}}
\label{fig:number3}
\end{figure}

The Nariai-type near-extremal condition states that the outer horizon is located close to the maximum of the function $\Delta_r$, as shown in Fig.\,\ref{fig:number3}\,(a). 
The near-extremal black hole still has three horizons: inner, outer, and cosmological. Moreover, the outer and cosmological horizons are close. When the Nariai-type extremal condition is satisfied, the two horizons are coincident and located at the maximum point of the function $\Delta_r$, as shown in Fig.\,\ref{fig:number3}\,(b). The final state of the Nariai-type near-extremal black hole can be specified based on the fluxes of a massive scalar field with an electric charge. The initial state is 
\begin{align}\label{eq:kerrinitial12}
\Delta_\text{max}\equiv \left.\Delta_r\right|_{r=r_\text{max}},\quad \Delta_\text{max}&\ll 1,\quad \left.\frac{\partial \Delta_r}{\partial r}\right|_{r=r_\text{max}}=\frac{\partial \Delta_\text{max}}{\partial r_\text{max}}=0,\quad \left.\frac{\partial^2 \Delta_r}{\partial r^2}\right|_{r=r_\text{max}}=\frac{\partial^2 \Delta_\text{max}}{\partial r_\text{max}^2}<0,
\end{align}
where $r_\text{max}$ is the location of the maximum, and its value is small and positive. Then, from Eq.\,(\ref{eq:Kerrminimum10}) to Eq.\,(\ref{eq:Kerrminimum20}), the change in the maximum value is
\begin{align}
\Delta_\text{max}+d\Delta_\text{max}&=\Delta_\text{max}+\frac{\partial \Delta_\text{max}}{\partial  M_\text{B}}dM_B+\frac{\partial \Delta_\text{max}}{\partial  J_\text{B}}dJ_\text{B}+\frac{\partial \Delta_\text{max}}{\partial  Q_\text{B}}dQ_\text{B}+\frac{\Delta_\text{max}}{\partial  r_\text{max}}dr_\text{max}\\
&=\Delta_\text{max}+\frac{4\pi}{\Xi}(\omega-m\Omega_\text{h}-q\Phi_\text{h})(\omega-m\Omega_\text{eff}'-q\Phi_\text{eff}')|\mathcal{T}|^2 \frac{\partial \Delta_\text{max}}{\partial M_\text{B}}dT,
\end{align}
where
\begin{align}
\Omega_\text{eff}'\equiv\left.{\frac{\partial \Delta_\text{max}}{\partial J_\text{B}}}\right/{\frac{\partial \Delta_\text{max}}{\partial M_\text{B}}},\quad \Phi_\text{eff}'\equiv \left.{\frac{\partial \Delta_\text{max}}{\partial J_\text{B}}}\right/{\frac{\partial \Delta_\text{max}}{\partial M_\text{B}}}.
\end{align}
To impose the near-extremal condition, we set $r_\text{max}-r_\text{h}=\epsilon\ll 1$. Hence, 
\begin{align}
\Omega_\text{eff}'&=-\Omega_\text{h}+\frac{2 a r_\text{h} \Xi}{(r_\text{h}^2+a^2)^2}\epsilon+\mathcal{O}(\epsilon^2),\quad 
\Phi_\text{eff}'=-\Phi_\text{h}+\frac{Q(r_\text{h}^2-a^2)}{(r_\text{h}^2+a^2)^2}\epsilon+\mathcal{O}(\epsilon^2),\\
{\frac{\partial \Delta_\text{max}}{\partial M_\text{B}}}&=-\frac{2(r_\text{h}^2+a^2)\Xi}{r_\text{h}}-2\Xi \left(1-\frac{a^2}{r_\text{h}^2}\right)\epsilon+\mathcal{O}(\epsilon^2),\nonumber\\
\Delta_\text{max}&=\frac{1}{3}(-3+a^2\Lambda + 6r_\text{h}^2\Lambda)\epsilon^2+\frac{8}{3}r_\text{h}\Lambda\epsilon^3+\Lambda \epsilon^4 + \mathcal{O}(\epsilon^5).\nonumber
\end{align}
The change in the maximum value is in its leading order, such that
\begin{align}
&\Delta_\text{max}+d\Delta_\text{max}=-\frac{8\pi|\mathcal{T}|^2(r_\text{h}^2+a^2) (\omega-m\Omega_\text{h}-q\Phi_\text{h})^2}{r_\text{h}}dT + \mathcal{O}(\epsilon).
\end{align}
This implies that the maximum value in the final state is {\it{negative}}, and that the change in this state is independent of the scalar field modes and asymptotic boundary condition with the square of the absolute transmission amplitude. Therefore, the final state is represented by the red line in Fig.\,\ref{fig:number3}\,(b). Here, no boundary exists between the black hole and cosmological horizon. Similar to the interior of a black hole, no static observer exists. Hence, it is not appropriate to discuss the WCC conjecture with regard to a static observer. Note that an observer can reach the inside of the inner horizon. However, this should be discussed in the context of the strong cosmic censorship conjecture for an infalling observer.

Instead of the WCC conjecture, we focus on the instability of the Nariai-type near-extremal black hole. Through a small perturbation of the scalar field, the interior of the black hole and the dS spacetime become connected\cite{Bousso:1999ms}. Thus, {\it{instability}} of the Nariai-type black hole arises owing to scattering of the scalar field. This instability is also independent of the scalar field modes. Hence, this outcome is classically inevitable. Note that this instability may potentially compete with a quantum effect in Nariai-type near-extremal black holes \cite{Gregory:2021ozs}.

\section{Laws of Thermodynamics}\label{sec6}

We have investigated the WCC conjecture for KN(A)dS black holes with arbitrary cosmological constant. The black holes were assumed to change under the fluxes carried by the scalar field in Eq.\,(\ref{eq:changeblackhole01}). As the changes in the black holes were obtained up to the first order, they should be consistent with the laws of thermodynamics, which the first orders of the thermodynamic variables satisfy. Thus, the changes in the fluxes can be confirmed as physical processes if they coincide with the laws of thermodynamics. Note that we consider an arbitrary cosmological constant. Thus, the initial states are non-extremal and Kerr-type extremal black holes. Here, we exclude Nariai-type near-extremal and extremal black holes because their final states are not black holes and because their thermodynamics are difficult to discuss.

The scalar field fluxes change the mass, angular momentum, and electric charge of the black hole. The outer horizon location also varies under the scattering. In particular, these changes can affect the surface area of the black hole, which is related to the Bekenstein-Hawking entropy in Eq.\,(\ref{eq:thermodynamicvar01}). The change in the entropy is
\begin{align}
dS_\text{h}=\frac{\partial S_\text{h}}{\partial  M_\text{B}}dM_B+\frac{\partial S_\text{h}}{\partial  J_\text{B}}dJ_\text{B}+\frac{\partial S_\text{h}}{\partial  Q_\text{B}}dQ_\text{B}+\frac{\partial S_\text{h}}{\partial  r_\text{h}}dr_\text{h},
\end{align}
where
\begin{align}
\frac{\partial S_\text{h}}{\partial  M_\text{B}}=&-\frac{2\pi((\Xi-1)r_\text{h}^2+a^2)}{M},\quad \frac{\partial S_\text{h}}{\partial  J_\text{B}}=-\frac{2\pi a ((r_\text{h}^2+a^2)\Lambda-3\Xi)}{3M},\\
\frac{\partial S_\text{h}}{\partial  Q_\text{B}}=&0,\quad \frac{\partial S_\text{h}}{\partial  r_\text{h}}=\frac{2\pi r_\text{h}}{\Xi},\quad d\Delta_\text{h}\equiv \left.d\Delta_r\right|_{r=r_\text{h}},\nonumber\\
dr_\text{h}=&\frac{2\Xi}{d\Delta_\text{h} M}(4Mr_\text{h}+2Q^2(\Xi-1)+a^2\Xi+r_\text{h}\Xi(r_\text{h}-3M-r_\text{h}\Xi))dM_\text{B}\nonumber\\
&+\frac{2a\Xi}{3d\Delta_\text{h} M}(r_\text{h}\Lambda(4M+r_\text{h}\Xi)-2Q^2\Lambda-3\Xi)dJ_\text{B}-\frac{2Q\Xi }{d\Delta_\text{h} M}dQ_\text{B}.\nonumber
\end{align}
Note that the horizon change can be positive or negative according to the scalar field mode. However, the change in entropy is finally obtained as
\begin{align}\label{eq:entropychange05}
dS_\text{h}= \frac{16\pi^2(r_\text{h}^2+a^2)(\omega-m\Omega_\text{h}-q\Phi_\text{h})^2|\mathcal{T}|^2}{\Xi d\Delta_\text{h}} dT,
\end{align}
where $d\Delta_\text{h}$ is positive and zero for non-extremal and extremal black holes, respectively. Thus, the change in the entropy is always positive, and the positive sign is independent of the scalar field mode and amplitude. Therefore, the scattering process follows the {\it{second law}} of thermodynamics. Furthermore, the change in the entropy of Kerr-type extremal black holes is divergent owing to the zero in the denominator in Eq.\,(\ref{eq:entropychange05}). This may be associated with the WCC conjecture and prevents the occurrence of a naked singularity due to the scattering. This aspect is clearer in relation to the third law of thermodynamics.

In accordance with the change in Eq.\,(\ref{eq:entropychange05}), we can construct the dispersion relation regarding the change in the mass. The change in the entropy can be rewritten as
\begin{align}\label{eq:thefirstlaw01}
d M_\text{B} = T_\text{h} d S_\text{h}+\Omega_\text{h}d J_\text{B} +\Phi_\text{h}dQ_\text{B}.
\end{align}
This is the {\it{first law}} of thermodynamics. Hence, without assuming the laws of thermodynamics, we can derive the first law. This outcome confirms that our flux-based analysis clearly shows physical processes and is consistent with the thermodynamics of the black hole. Moreover, the derivation of Eq.\,(\ref{eq:thefirstlaw01}) is independent of any particular boundary condition or cosmological constant.

\begin{figure}[!h]
\centering
\subfigure[{$(1,0,0)$ mode in $(0,1)$ black hole.}]
{\includegraphics[scale=0.38,keepaspectratio]{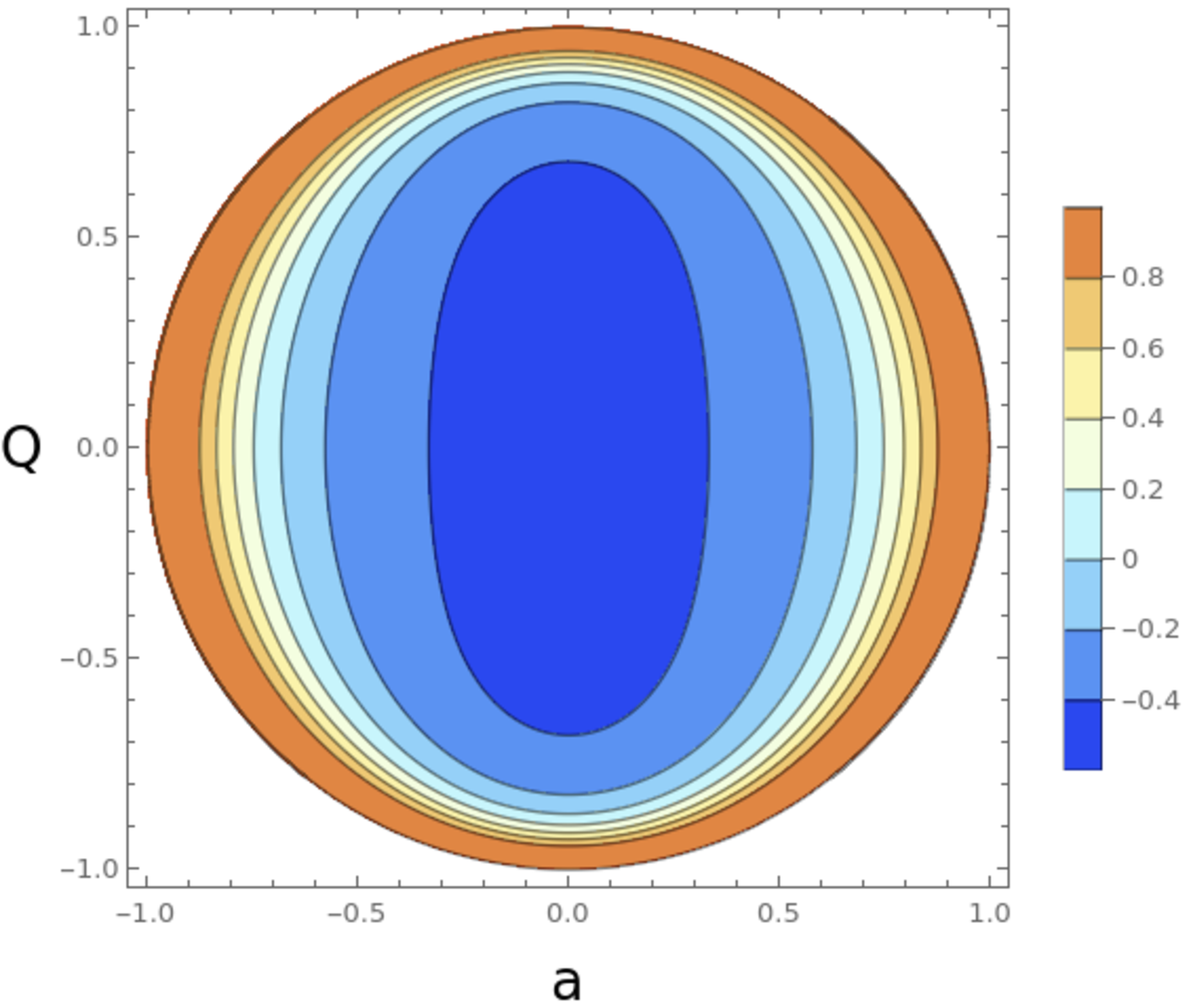}}
\subfigure[{$(1,1,0)$ mode in $(0,1)$ black hole.}]
{\includegraphics[scale=0.38,keepaspectratio]{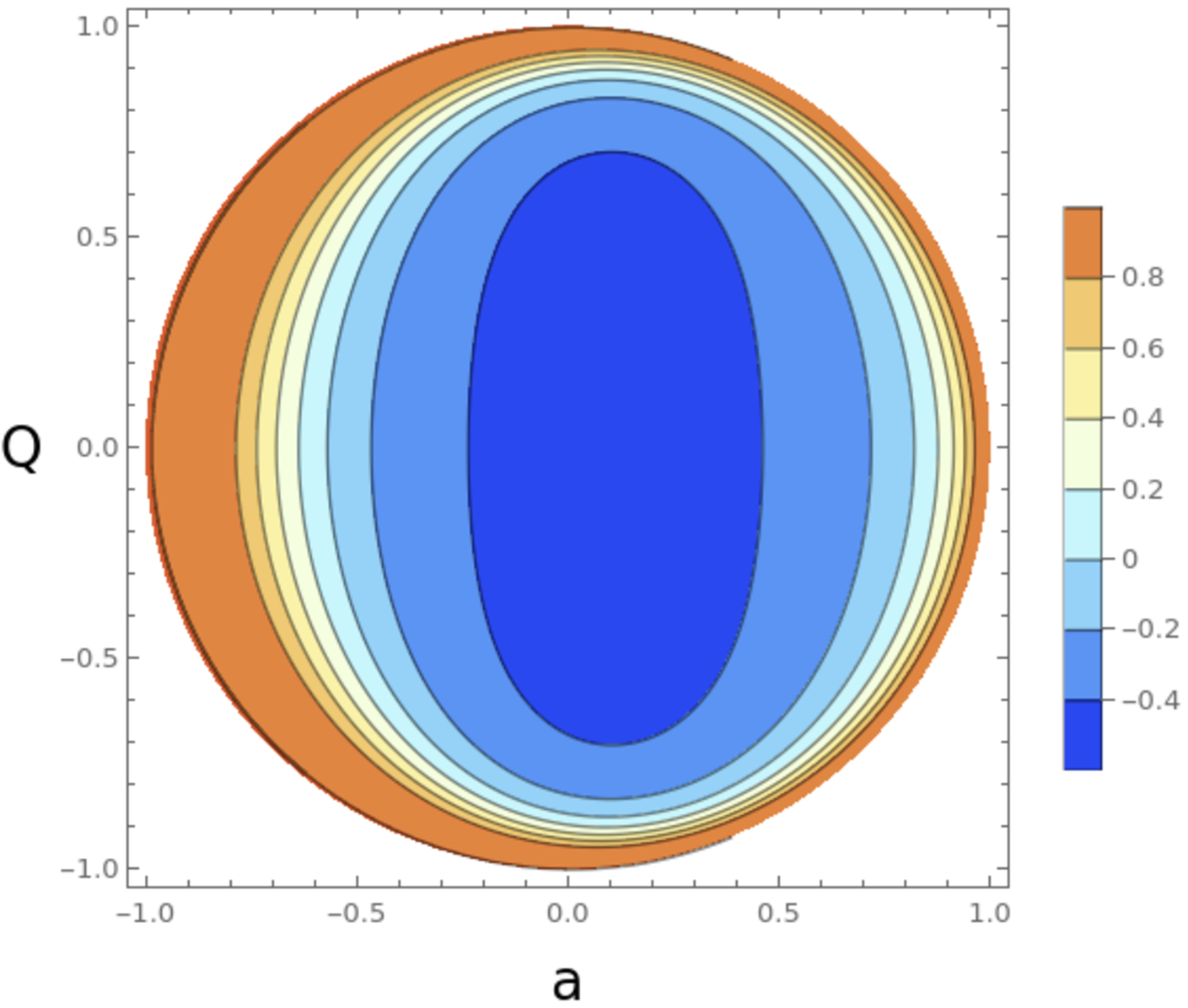}}
\subfigure[{$(1,1,1)$ mode in $(0,1)$ black hole.}]
{\includegraphics[scale=0.38,keepaspectratio]{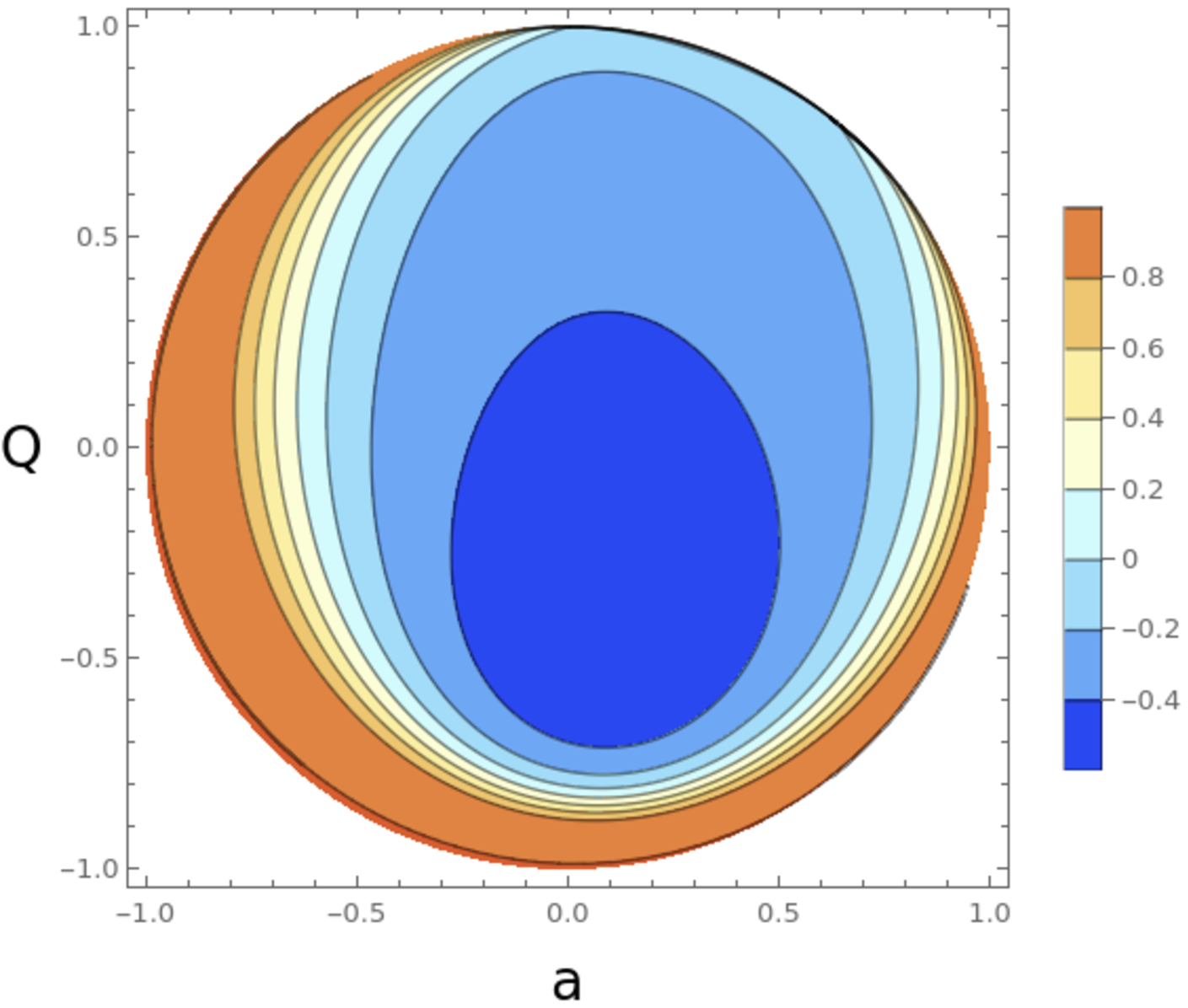}}
\subfigure[{$(1,0,0)$ mode in $(-1,1)$ black hole.}]
{\includegraphics[scale=0.38,keepaspectratio]{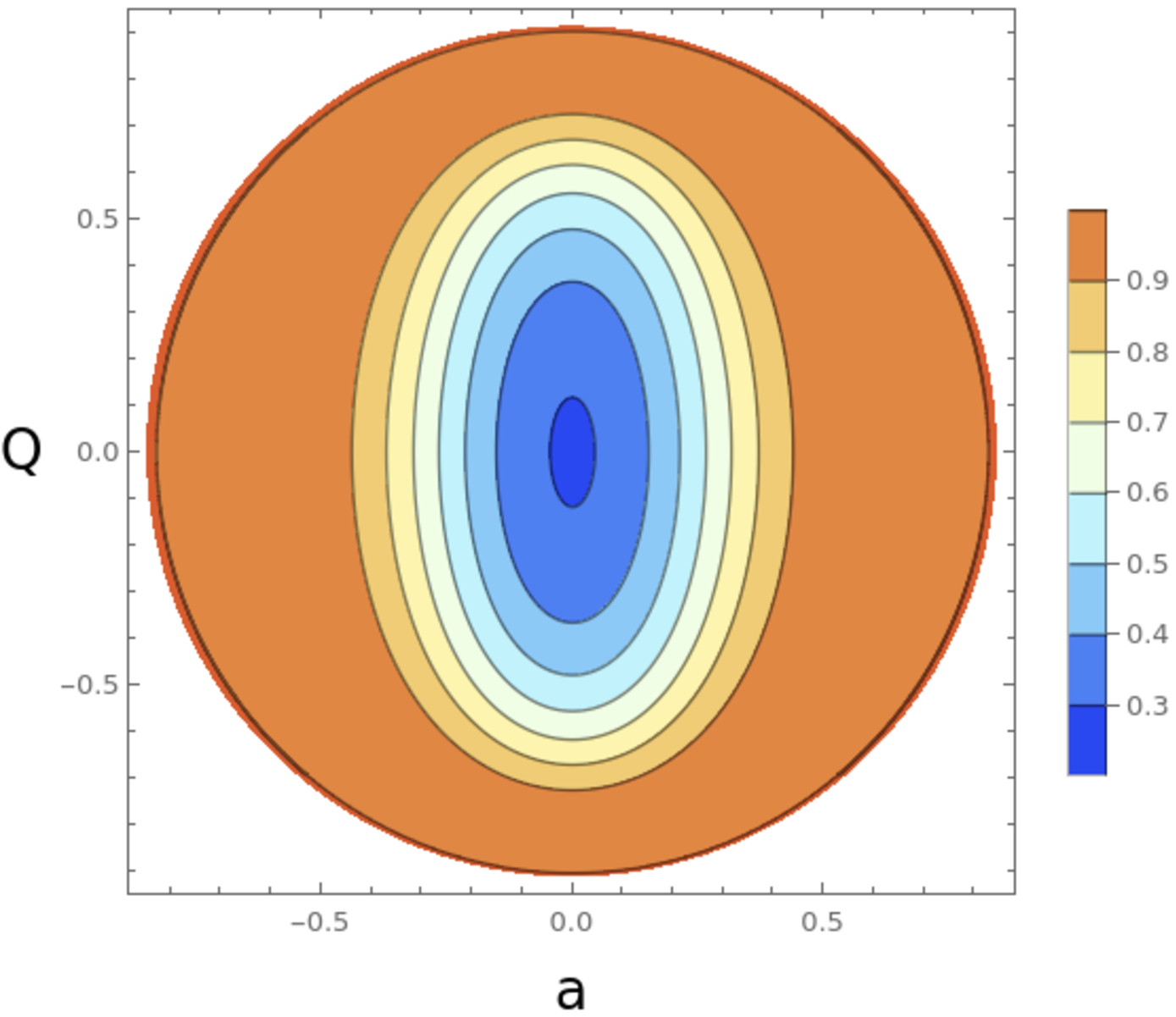}}
\subfigure[{$(2,1,1)$ mode in $(-1,1)$ black hole.}]
{\includegraphics[scale=0.38,keepaspectratio]{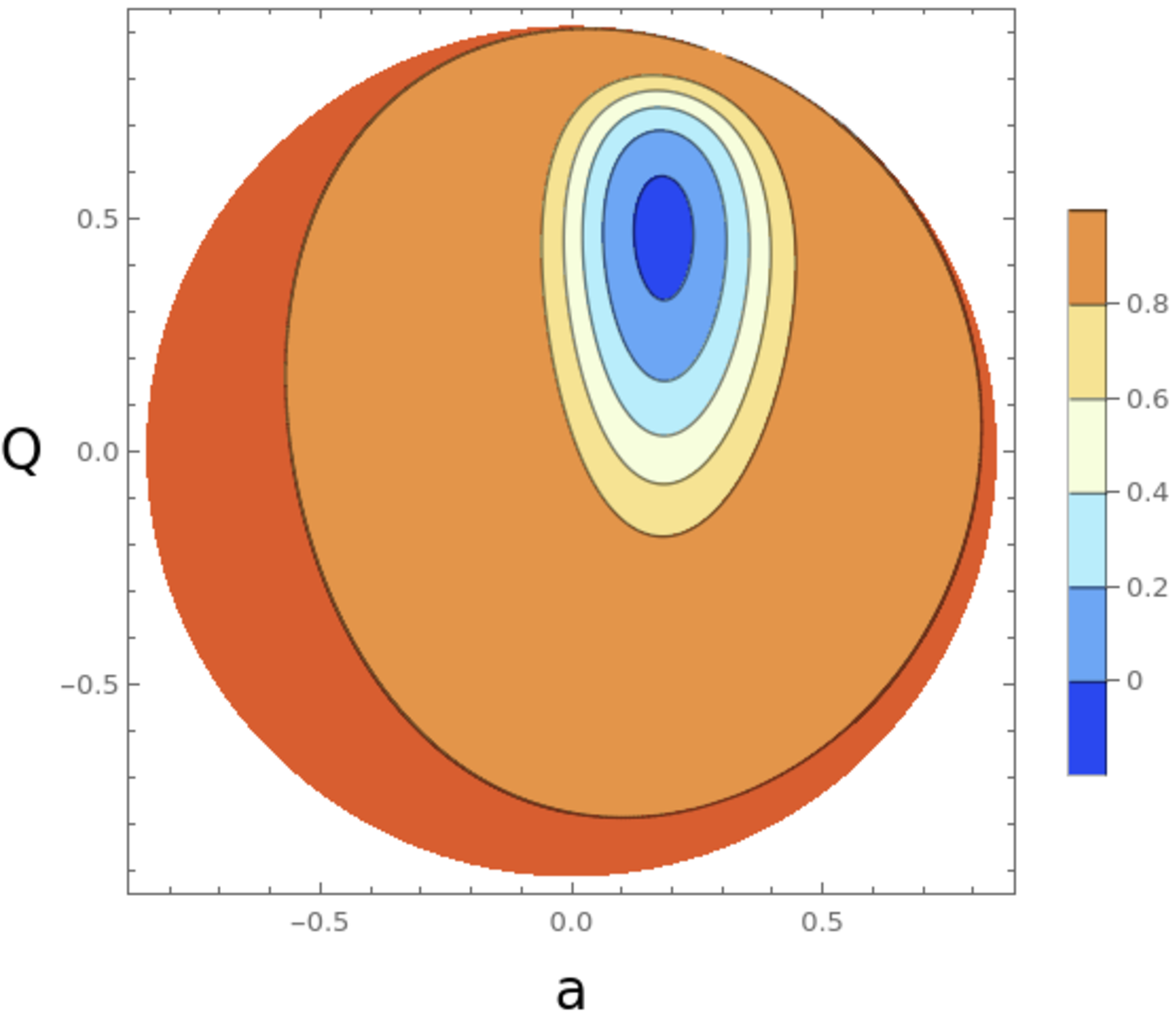}}
\subfigure[{$(2,1,2)$ mode in $(-1,1)$ black hole.}]
{\includegraphics[scale=0.38,keepaspectratio]{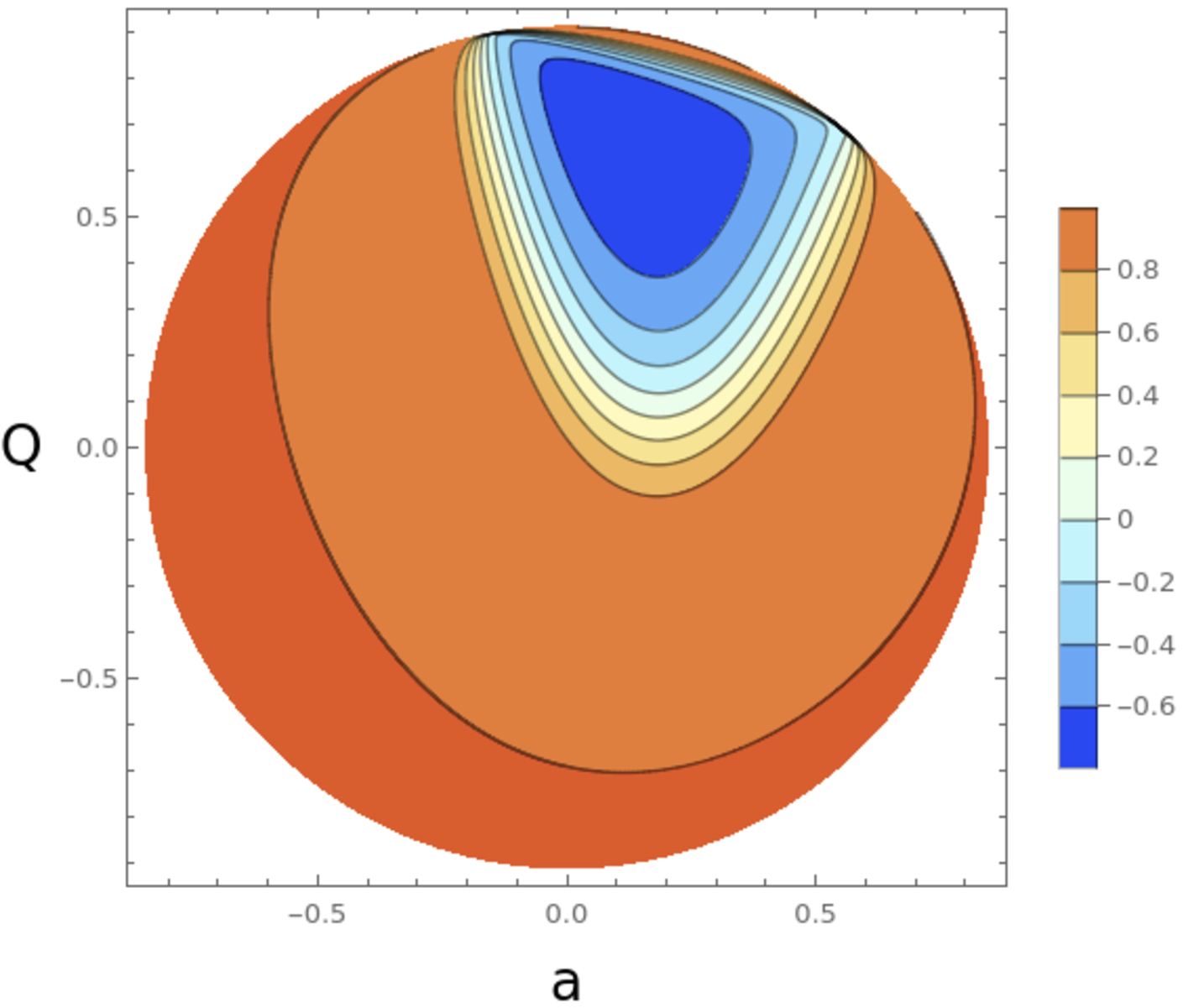}}
\subfigure[{$(1,0,0)$ mode in $(0.1,1)$ black hole.}]
{\includegraphics[scale=0.38,keepaspectratio]{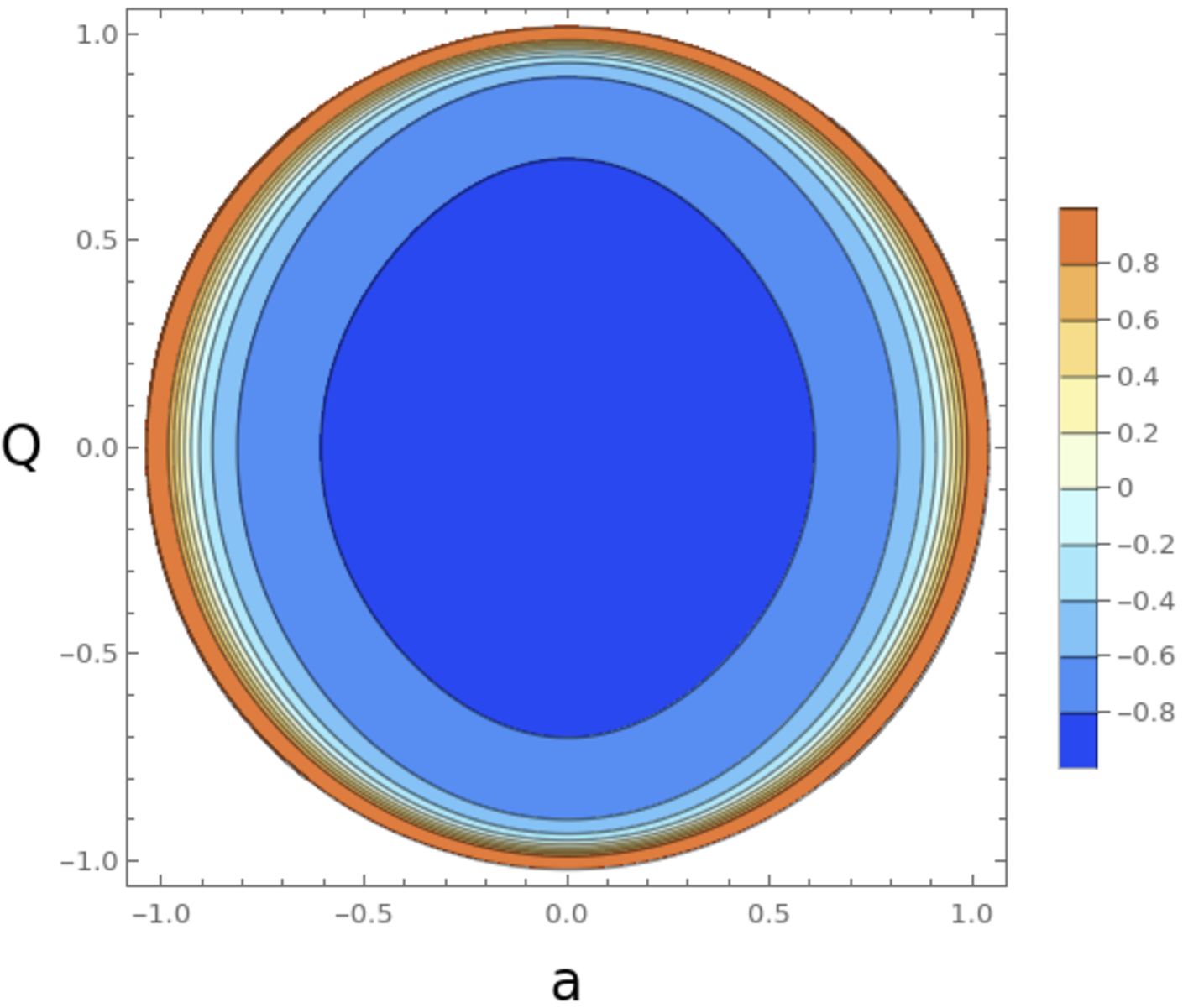}}
\subfigure[{$(1,1,1)$ mode in $(0.1,1)$ black hole.}]
{\includegraphics[scale=0.38,keepaspectratio]{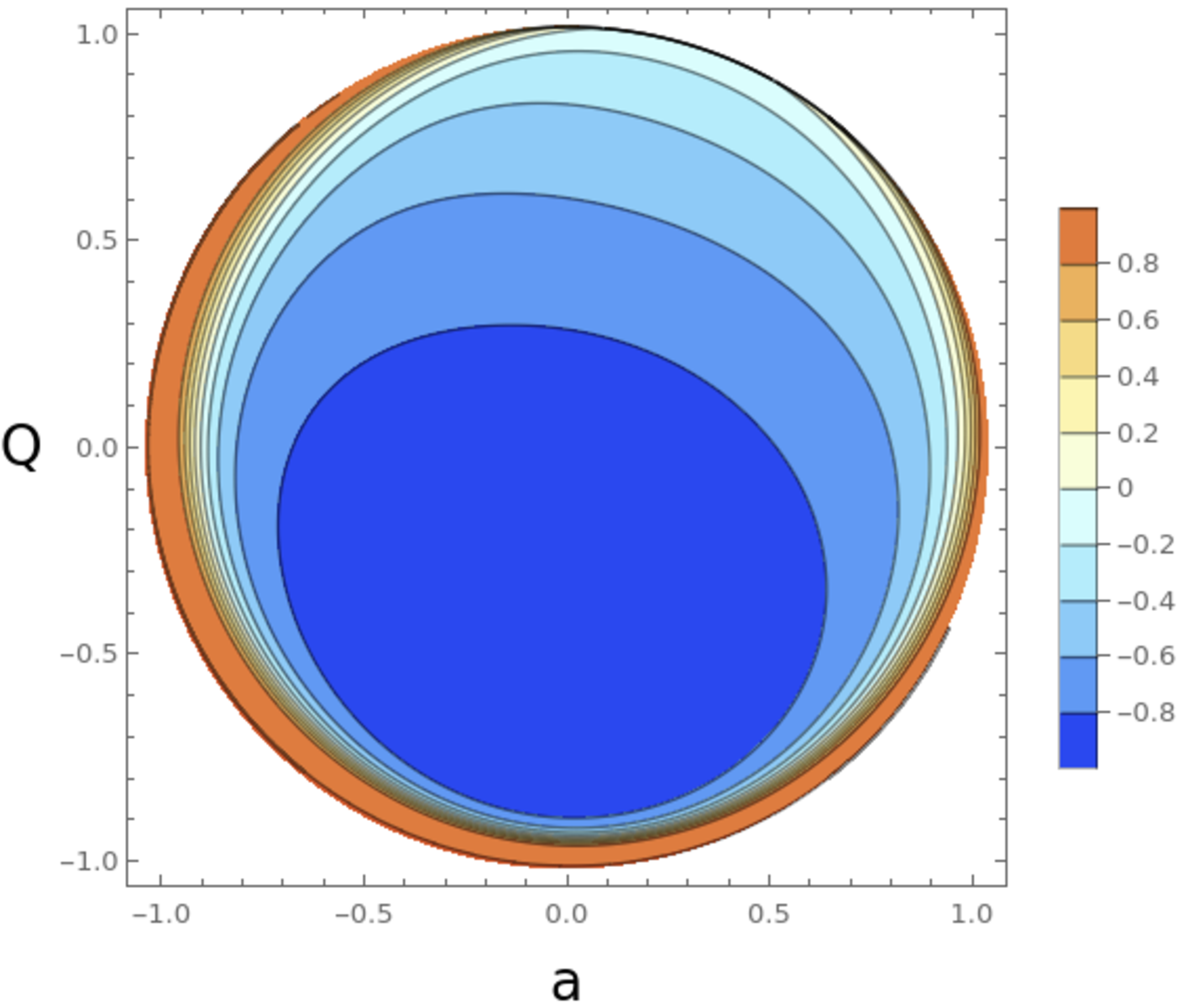}}
\subfigure[{$(2,2,3)$ mode in $(0.1,1)$ black hole.}]
{\includegraphics[scale=0.38,keepaspectratio]{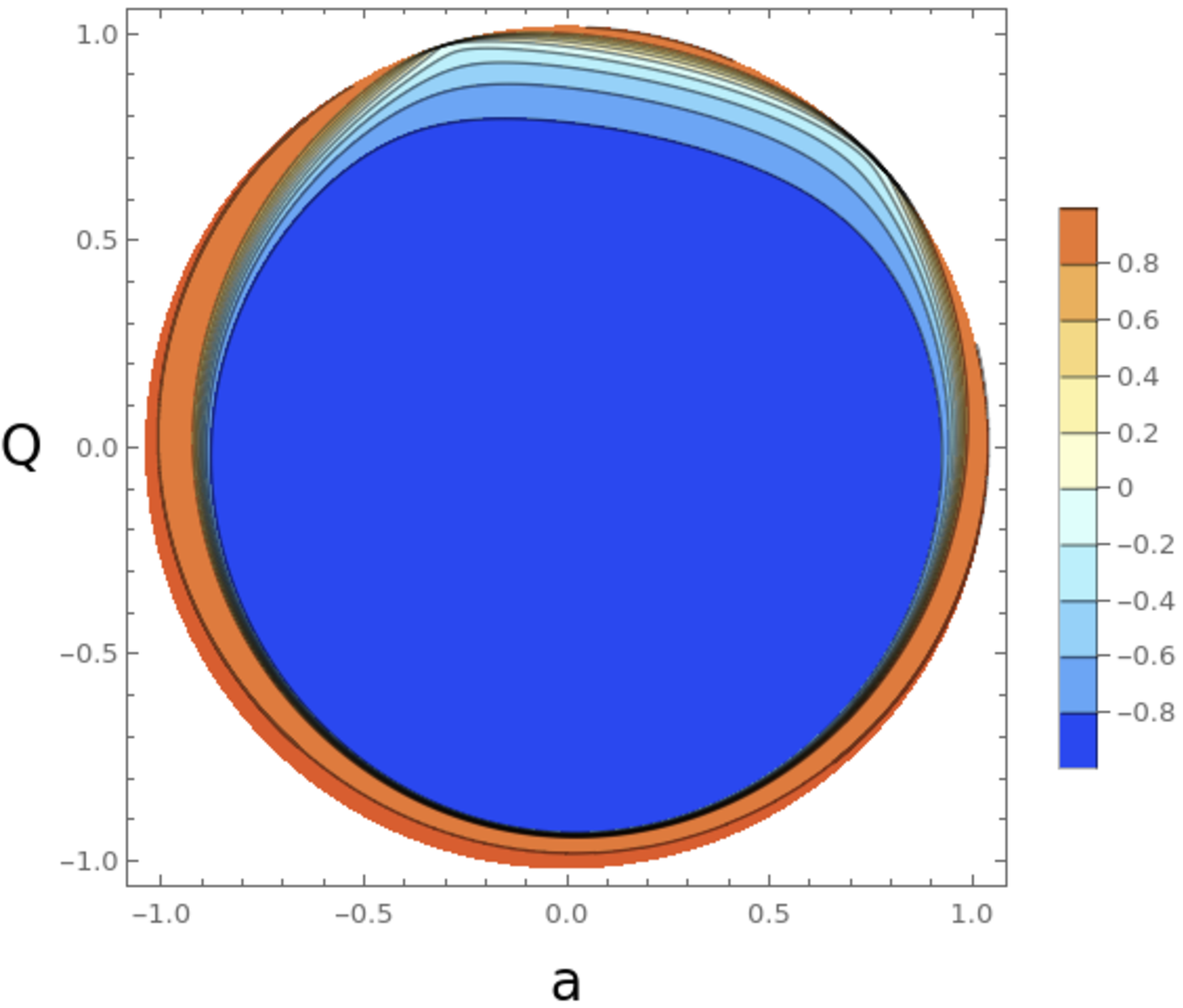}}
\caption{{\small Diagrams of $\tanh\left(\frac{d T_\text{h}}{dT}\right)$ in $(\omega,m,q)$ modes in $(\Lambda, M)$ black holes.}}
\label{fig:number4}
\end{figure}

In this work, the main initial states considered are the Kerr-type near-extremal and extremal black holes. In thermodynamics, such extremal black holes correspond to an interesting state of zero Hawking temperature. Thermodynamically, zero temperature is associated with the third law of thermodynamics, which is expressed in various ways. In accordance with the well-known version \cite{Bardeen:1973gs}, we herein consider the third law, which states that zero temperature cannot be attained in a black hole through finite physical processes. The change in the Hawking temperature owing to scalar field scattering is
\begin{align}
dT_\text{h}=\frac{\partial T_\text{h}}{\partial M_\text{B}}dM_\text{B}+\frac{\partial T_\text{h}}{\partial J_\text{B}}dJ_\text{B}+\frac{\partial T_\text{h}}{\partial Q_\text{B}}dQ_\text{B}+\frac{\partial T_\text{h}}{\partial r_\text{h}}dr_\text{h},
\end{align}
where
\begin{align}
\frac{\partial T_\text{h}}{\partial M_\text{B}}&=\frac{a^2 r_\text{h}\left(1-\frac{a^2+Q^2}{r_\text{h}^2}-\frac{a^2 \Lambda}{3}-r_\text{h}^2\Lambda\right)\Xi^2}{2M\pi(r_\text{h}^2+a^2)^2}+\frac{r_\text{h}\left(\frac{2a^2\Lambda \Xi^2}{3M}+\frac{\frac{4a^2Q^2\Lambda\Xi}{3M}+\frac{2a^2\Xi^2}{M}}{r_\text{h}^2}\right)}{4\pi (r_\text{h}^2+a^2)},\\
\frac{\partial T_\text{h}}{\partial J_\text{B}}&=-\frac{a r_\text{h}\left(1-\frac{a^2+Q^2}{r_\text{h}^2}-\frac{a^2 \Lambda}{3}-r_\text{h}^2\Lambda\right)\Xi^2}{2M\pi(r_\text{h}^2+a^2)^2}-\frac{r_\text{h}\left(\frac{2a\Lambda \Xi^2}{3M}+\frac{\frac{4aQ^2\Lambda\Xi}{3M}+\frac{2a\Xi^2}{M}}{r_\text{h}^2}\right)}{4\pi (r_\text{h}^2+a^2)},\quad \frac{\partial T_\text{h}}{\partial Q_\text{B}}=-\frac{Q\Xi}{2\pi r_\text{h}(r_\text{h}^2+a^2)},\nonumber\\
\frac{\partial T_\text{h}}{\partial r_\text{h}}&=\frac{r_\text{h}\left(\frac{2(a^2+Q^2)}{r_\text{h}^3}-2r_\text{h} \Lambda\right)}{4\pi(r_\text{h}^2+a^2)}-\frac{r_\text{h}^2\left(1-\frac{a^2+Q^2}{r_\text{h}^2}-\frac{a^2 \Lambda}{3}-r_\text{h}^2\Lambda\right)\Xi^2}{2\pi(r_\text{h}^2+a^2)^2}+\frac{1-\frac{a^2+Q^2}{r_\text{h}^2}-\frac{a^2 \Lambda}{3}-r_\text{h}^2\Lambda}{4\pi(r_\text{h}^2+a^2)}.\nonumber
\end{align}
This provides the general form of the change in all KM(A)dS black holes. However, it is extremely complicated. Furthermore, the change in temperature depends on the scalar field mode. Hence, this change has no particular direction, as shown in Fig.\,\ref{fig:number4}. However, if the initial state is assumed to correspond to a Kerr-type near-extremal black hole, the change in temperature becomes positive for any cosmological constant.
\begin{align}\label{eq:thethirdlaw01}
dT_\text{h}=\frac{2|\mathcal{T}|^2(\omega-m\Omega_\text{h}-q\Phi_\text{h})^2}{ r_\text{h}\epsilon}dT+\mathcal{O}(\epsilon^0),
\end{align}
where $\epsilon$ is the difference between the locations of the minimum and outer horizon in Eq.\,(\ref{eq:epsilonandminimum}). When the initial state is close to the extremal condition, the difference $\epsilon$ becomes small. Then, the temperature increases significantly along with the first term in Eq.\,(\ref{eq:thethirdlaw01}). This is also shown in Fig.\,\ref{fig:number4}, where the changes in temperature are always positive for a near-extremal black hole. Note that we should consider the minimum condition in Eq.\,(\ref{eq:kerrinitial01}) in the context of Eq.\,(\ref{eq:thethirdlaw01}) to exclude the Nariai-type extremal condition in Eq.\,(\ref{eq:kerrinitial12}). Therefore, for the initial near-extremal black hole with $\epsilon\ll1$, the temperature increases in the scattering process and, thus, it is impossible for zero temperature to be attained in the extremal black hole. This is the {\it{third law}} of thermodynamics.

From a thermodynamic perspective, for an arbitrary cosmological constant, non-extremal black holes cannot become Kerr-type extremal black holes through scalar field scattering, because the temperature of the near-extremal black hole increases substantially, as shown in Eq.\,(\ref{eq:thethirdlaw01}). This is ensured by the third law of thermodynamics. Moreover, with the scattering, the Bekenstein-Hawking entropy in the near-extremal black hole also increases considerably, as shown in Eq.\,(\ref{eq:entropychange05}). In other words, an increase in temperature is thermodynamically preferred under the second law of thermodynamics. The first law of thermodynamics ensures that the scattering process is physical. Again, if overcharging results from the scattering, the initial non-extremal black hole successively becomes an extremal black hole and a naked singularity in finite operations. However, at the near-extremal black hole, the temperature increases and, thus, the state moves far from the extremal black hole. Thus, the black hole cannot be overcharged. This outcome is also thermodynamically preferred. Therefore, the validity of the WCC conjecture is also supported by the laws of thermodynamics.

\section{Summary}\label{sec7}

We investigated the WCC conjecture for the KN(A)dS black hole with an arbitrary cosmological constant $\Lambda$. The KN(A)dS black hole has three conserved quantities: mass, angular momentum, and electric charge. Consideration of an arbitrary cosmological constant also enabled investigation of the asymptotically flat AdS and dS geometries. Hence, based on our study of the KN(A)dS black hole, we can arrive at the generalized conclusion that the WCC conjecture is valid for black holes.

To validate the WCC conjecture, the changes in a Kerr-type near-extremal KN(A)dS black hole owing to scattering from a massive scalar field were considered. To determine the correct scalar field fluxes, we imposed a coordinate transformation on the KN(A)dS black hole metric in Boyer-Lindquist coordinates to obtain an asymptotically static boundary for nonzero cosmological constants. Then, we assumed that the conserved quantities of the black hole change with the scalar field fluxes at the outer horizon, such that the initial state $(M_\text{B},J_\text{B},Q_\text{B})$ becomes the final state $(M_\text{B}+dM_\text{B},J_\text{B}+dJ_\text{B},Q_\text{B}+dQ_\text{B})$ during an infinitesimal time interval. By examining the behaviors of the function $\Delta_r$, we found that the outer horizon $r_\text{h}$ exists stably under the scattering. Hence, the WCC conjecture is valid. In particular, the horizon stability is ensured for any scalar field mode with any asymptotic boundary condition. The asymptotic boundary condition must determine the scalar field transmission amplitude at the outer horizon. However, the amplitude is introduced as the square of the absolute value of the variation in the minimum value of $\Delta_r$. Therefore, the transmission amplitude always makes a positive contribution, which ensures a stable outer horizon for any asymptotic geometry.

Moreover, we found that a Nariai-type near-extremal black hole with a positive cosmological constant exhibits instability. Although the scalar field fluxes are infinitesimally small, the outer and cosmological horizons can become connected and form a spacetime similar to the interior of a black hole under the considered scattering. This result shows that the Nariai-type near-extremal black hole is unstable. In this final state, the outside observer cannot be defined. Therefore, this state is far from the WCC conjecture. However, the mechanism is essentially identical to that employed for the Kerr-type near-extremal black hole.

Our conclusion regarding the validity of the WCC conjecture could also be associated with the laws of thermodynamics. The first law was obtained by rewriting the fluxes at the outer horizon. Since our analysis was based on the flux-induced changes, this outcome confirmed that we had considered a thermodynamically consistent physical process. By confirming the WCC conjecture, we found that the Kerr-type near-extremal black hole becomes a non-extremal black hole because the minimum of the function $\Delta_r$ moves in the negative direction. The second law confirmed the WCC conjecture and showed that such a change is thermodynamically preferred based on the steep increase in entropy. We also confirmed the third law, which indicates that the finite operations of a physical process cannot form an extremal black hole with zero temperature. This finding is consistent with Kerr-type near-extremal black holes becoming non-extremal black holes, according to our results. Therefore, the laws of thermodynamics also validate the WCC conjecture and, hence, the WCC conjecture is safe.

\vspace{10pt} 

\noindent{\bf Acknowledgments}

\noindent This work was supported by the National Research Foundation of Korea (NRF) grant funded by the Korea government (MSIT) (NRF-2018R1C1B6004349) and the Dongguk University Research Fund of 2021. BG appreciates APCTP for its hospitality during the topical research program, {\it Gravity and Cosmology}.

\end{document}